\documentstyle[aps,preprint]{revtex}

\begin{document}
\title{Damping of Very Soft Moving Quarks in High-Temperature QCD}
\author{A. Abada$^{1}\thanks{{\tt abada@wissal.dz}}$, K. Bouakaz$^{1}$ and N.
Daira-Aifa$^{2}$}
\address{$^{1}$D\'{e}partement de Physique, Ecole Normale Sup\'{e}rieure\\
BP 92 Vieux Kouba, 16050 Alger, Algeria\smallskip\\
$^{2}$Facult\'{e} de Physique,\\
\ Universit\'{e} des Sciences et de la Technologie Houari Boumediene\\
BP\ 32 El-Alia, Bab Ezzouar 16111 Alger, Algeria\smallskip }
\date{\today}
\maketitle

\begin{abstract}
We determine the analytic expression of the damping rates for very soft
moving quarks in an expansion to second order in powers of their momentum in
the context of QCD\ at high temperature. The calculation is performed using
the hard-thermal-loop-summed perturbation scheme. We describe the range of
validity of the expansion and make a comparison with other calculations,
particularly those using a magnetic mass as a shield from infrared
sensitivity. We discuss the possible occurrence of infrared divergences in
our results and argue that they are due to magnetic sensitivity.
\end{abstract}

\bigskip

{\small pacs: 11.10.Wx 12.38.-t 12.38.Bx 12.38.Mh}

{\small keywords: hard thermal loops. soft quark damping. infrared
sensitivity.}

ENSK--TP--18

\vspace{0.3in}

\section{Introduction}

One runs into difficulties when one applies standard loop expansion to gauge
theories at high temperature $T$: physical quantities like the dispersion
laws become gauge dependent. Early work on the QED plasma using a
hydrodynamic approach is \cite{silin}, followed by \cite{tsytovich}, taking
account of one-loop quantum effects. The work \cite{kalash-klimo} discusses
to one-loop order the QCD polarization tensor at high temperature and quark
density and determines the gluon dispersion laws to lowest order in the
coupling $g$. It shows that these dispersion laws are gauge invariant but
the one-loop-order gluon damping rates in the long-wavelength limit are not.
It also shows that while chromoelectric Debye screening does occur to lowest
order in the one-loop calculation, chromomagnetic screening does not, a
gauge invariant statement. The non-screening of chromomagnetic fields at
lowest order is also discussed in \cite{linde1,linde2,gross-pisarski-yaffe}.
The massless-quark spectrum to lowest order in the coupling is described in 
\cite{klimov1} and the full quasiparticle spectra to lowest order at high $T$
for the whole momentum range are given in \cite{klimov2}. The quasiparticle
spectra are also described in \cite{weldon1} for gluons and \cite{weldon2}
for quarks.

The problem of gauge dependence of the damping rates has been emphasized in
works in which the gluon damping rates, particularly at zero momentum, have
been calculated to one-loop order in various gauges and schemes and
different results have been obtained \cite{later}. It was realized that the
problem is related to the fact that at high temperature, higher-loop
diagrams can contribute to lower orders in powers of the coupling \cite
{pisarski1}. In other words, the standard loop expansion is no more an
expansion in powers of $g^{2}$. In a series of papers, Braaten and Pisarski
developed a systematic method for an effective perturbative expansion that
sums the so-called hard thermal loops (HTL)\ into effective propagators and,
equally important for gauge theories, effective vertices \cite
{BPhtl,frenkel-taylor,le bellac}. Using this method, the transverse-gluon
damping rate $\gamma _{t}(0)$ at zero momentum was shown to be finite,
positive and independent of the gauge \cite{BPgamt}. Later, a
generating-functional formalism for high-$T$ QCD in the HTL approximation
was developed \cite{action} and a relation to the eikonal of a Chern-Simons
gauge theory was found \cite{Chern-Simons}. From there, a hydrodynamic
approach showed that the HTL approximation is essentially `classical' \cite
{hydro}.

Once developed, the important question to answer is whether the HTL-summed
perturbation is reliable for calculations in QCD\ at high temperature. If
so, it would constitute an adequate framework for describing the properties
of the quark-gluon plasma. Of particular interest is the question of
infrared sensitivity in massless gauge theories, worsened at finite
temperature by the presence of the Bose-Einstein distribution which behaves
like $1/k$ for very small gluon energies $k$: quantities tend to diverge
like powers of the infrared cutoff rather than logarithmically as is the
case at zero temperature \cite{burgess-marini--rebhan}. This infrared
problem is prior to the advent of the HTL scheme. It is for example shown in 
\cite{baier-pire-schiff--altherr-aurenche-becherrawy} that infrared (and
mass) singularities do occur but cancel out in first-order radiative
corrections to the production of lepton pairs in thermal (massless) QCD. It
is therefore most interesting to see if the HTL scheme is of any help in
this regard. More precisely, does the HTL-summed perturbation constitute a
workable framework in which infrared divergences are cured consistently
order by order in the coupling?

It turns out that the HTL summation dresses the massless quarks and gluons
allowing them to acquire thermal masses of order $gT$, $m_{f}$ and $m_{g}$
respectively \cite{kalash-klimo,klimov1,klimov2,weldon1,weldon2}. This means
that to this lowest order $gT$ in effective perturbation, the infrared
region is `safe'. But as recalled, static chromomagnetic fields do not
screen at this lowest order, they are believed to do so at the
next-to-leading order $g^{2}T$, the so-called magnetic scale \cite
{linde1,linde2,gross-pisarski-yaffe}. Therefore, the test for the infrared
safeness of the HTL-summed expansion starts really in next-to-leading order
calculations.

Much work has been carried regarding the use of the HTL summation up to
next-to-leading order. For example, it is shown in \cite
{baier-nakkagawa-niegawa-redlich} that the HTL\ summation cures the infrared
logarithmic divergence in the production rate of hard thermal photons in
high-temperature QCD with massless quarks. Also, the damping of fast moving
fermions has been thoroughly investigated in \cite
{pisarski1,lebedev-smilga,burgess-marini--rebhan,braaten-thoma,thoma-gyulassy}
for hot QCD\ and \cite{altherr-petitgirard-del rio gaztelurrutia,baier-kobes}
for hot QED. In \cite{altherr-petitgirard-del rio gaztelurrutia} for
instance, it is first performed a bare (i.e., not HTL-summed) two-loop
calculation and by taking the fermion slightly off-shell, there occurs to
this order cancellation of infrared, both electric and magnetic,
singularities. It is then shown that the dominant on-shell graphs are those
dictated by the HTL summation. Also, it is shown in \cite
{burgess-marini--rebhan} that it is in fact possible to find the leading
contribution $g^{2}T\ln 1/g$ to the damping rates for energetic (hard)
quarks and gluons in high-temperature QCD without having recourse to the
full machinery of the HTL-summation program, a leading contribution already
obtained for quarks in \cite{pisarski1}, see also{\footnotesize \ }\cite
{lebedev-smilga}.

It seems therefore that the infrared problem can somewhat be brought under
control when it comes to describing fast-moving (hard) quasiparticles. But
how about slow-moving (soft) quarks and gluons, particularly those on shell?
A first indication of the sensitivity of the HTL perturbation to slow moving
particles can be found in \cite
{baier-peigne-schiff--aurenche-becherrawy-petitgirard} where the production
of non-thermalized soft real photons in HTL-summed perturbation in high-$T$
QCD is discussed. It is argued that this scheme fails to screen mass
singularities in that it is not able to yield a finite contribution to
leading order to the production rate, a physical quantity. However, the
divergences involved in \cite
{baier-peigne-schiff--aurenche-becherrawy-petitgirard} are collinear in
nature and come from dressed vertices. In this regard, an improved action
has been proposed in \cite{flechsig-rebhan} which incorporates an asymptotic
mass $m_{\infty }$ that removes singularities coming from light-like
external momenta.

As to the importance of the magnetic sector and the infrared sensitivity of
the HTL\ scheme at next-to-leading order calculations, this is well
demonstrated in the works \cite{rebhan1,braaten-nieto1,rebhan2}. Indeed, 
\cite{rebhan1} calculates in HTL-summed perturbation the non-abelian Debye
mass at next-to-leading order from the static limit of the polarization
tensor. \cite{braaten-nieto1} determines the same physical quantity at the
same order in the same scheme, but from the correlator of two Polyakov
loops, a gauge invariant quantity. Paper \cite{rebhan2} discusses the more
general problem of next-to-leading order non-abelian Debye screening in
one-loop HTL-summed perturbation. It argues that since the magnetic sector
is nonperturbative in essence, the perturbative next-to-leading-order
results may not be reliable. This is explicitly shown in, for example, the
strong dependence of the analytic structure of the inverse of the static
longitudinal propagator on the infrared behavior of the transverse gluons
where the results differ significantly depending on whether we regularize
the infrared sector by introducing a magnetic mass\footnote{%
Introduced as an infrared regulator, a point we come to later in section
three.} or not. One other interesting point discussed is the important role
of the magnetic mass in cancelling the gauge dependent terms when obtaining
Debye screening from the Polyakov-loop correlator. Finally, comparison with
lattice simulations indicates that the magnetic-mass enhanced results are
more compatible with the lattice ones, hence the importance of the magnetic
sector.

However, Debye screening is static in nature. It is therefore interesting to
examine dynamic on-shell quantities in order to understand better the
infrared behavior of hot gauge theories in HTL-summed perturbation for soft
moving quasiparticles. It turns out that the damping rates at lowest order
for such (very) soft moving quasiparticles are quite suitable. Indeed, the
HTL quasiparticle self-energies are real and so no damping is manifest at
order $gT$; it starts at precisely the magnetic scale $g^{2}T$. Thus, to
exhibit damping, one needs to add to the inverse propagators the
next-to-leading order contributions to the self-energies which are, {\it as
dictated} by the HTL-summed expansion, one-loop corrections with soft loop
momenta, hence all propagators and vertices have to be HTL dressed.

The first such calculation is paper \cite{BPgamt} which determined the
damping rate $\gamma _{t}(0)$ for transverse on-shell gluons with zero
momentum and found: 
\begin{equation}
\gamma _{t}(0)=0.088N_{c}\,g^{2}T,  \label{gamt-zero}
\end{equation}
where $N_{c}$ is the number of colors. The analytic calculation of $\gamma
_{t}(p)$ to order $\left( p/m_{g}\right) ^{2}$ where $p$ is the momentum of
the very soft gluon was carried in \cite{AAT}. The zeroth order (\ref
{gamt-zero}) was recovered and it was indicated that the coefficient of the
order $\left( p/m_{g}\right) ^{2}$ may carry infrared divergences. The
infrared sensitivity of the on-shell gluon damping rates has been emphasized
in \cite{AAB,AA} where the damping rate $\gamma _{l}(0)$ for longitudinal
gluons with zero momentum was determined to lowest order $g^{2}T$ and found
to be different from $\gamma _{t}(0)$ and infrared divergent. This is to be
contrasted with the fact that at zero momentum, there must be no difference
between longitudinal and transverse gluons \cite{BPgamt}. This statement is
emphasized in \cite{dirks-niegawa-okano} where a Slavnov-Taylor identity for
the gluon polarization tensor in Coulomb gauge is derived and when applied
to the next-to-leading order gluon self-energy, the equality $\gamma
_{l}(0)=\gamma _{t}(0)$ is obtained\footnote{%
It is assumed in \cite{dirks-niegawa-okano} that the spatial next-to-leading
order HTL-summed gluon self-energy is isotropic at zero momentum. Our
explicit and direct calculations do not recover this isotropy. This issue
will be addressed in detail in \cite{AABo}.}.

The next step must be the discussion of the damping rates for very soft
moving quarks at lowest order $g^{2}T$. This is because quarks are also
important in the structure of hot QCD. Since they too acquire a thermal mass 
$m_{f}$ at order $gT$ and their damping rates start at the magnetic scale $%
g^{2}T$, it is all but legitimate to inquire about their infrared
sensitivity. There are already the two works \cite{kobes-kunstatter-mak} and 
\cite{braaten-pisarski (quarks)} which determined independently the damping
rates $\gamma _{\pm }(0)$ for quarks with zero-momentum and found: 
\begin{equation}
\gamma _{\pm }(0)=a_{0}C_{f}\,g^{2}T,  \label{gamq-zero}
\end{equation}
where $C_{f}=(N_{c}^{2}-1)/2N_{c}$ and $a_{0}$ is a finite constant
depending on $N_{c}$ and $N_{f}$, the number of flavors. For example, for $%
N_{c}=3$ and $N_{f}=2$, we have $a_{0}=0.111\dots $, \cite
{kobes-kunstatter-mak,braaten-pisarski (quarks)}. Result (\ref{gamq-zero})
for quarks resembles result (\ref{gamt-zero}) for gluons. It is therefore
interesting to determine the damping rates for {\it moving} quarks, but with 
{\it very} soft momenta. In this article, we attempt to obtain to lowest
order $g^{2}T$ an analytic expression for the damping rates $\gamma _{\pm
}(p),$ where ${\bf p}=p\widehat{{\bf p}}$ is the momentum of the quark,
using {\it only} as ingredients what the HTL-summed expansion dictates.

Obtaining a compact expression for $\gamma _{\pm }(p)$ would be ideal but
hardly feasible technically. Rather, we attempt to obtain an expression for $%
\gamma _{\pm }(p)$ in powers of $p/m_{f}$ up to second order. This expansion
is carried early on in the calculation. We will specify in section three its
range of validity and argue that in order to get an explicit expression for
the damping rates, manipulating otherwise is practically intractable. In
this work, we describe in detail how we obtain the analytic expressions for
the first three coefficients $a_{0}$, $a_{1}$ and $a_{2}$ involved in the
expansion, see (\ref{gam-pm-final}), and we defer the numerical evaluation
of these to future work \cite{ABD}. This is because it necessitates the
extraction of the potentially infrared-divergent pieces from the finite
contributions, something somewhat complicated. An additional complication
comes from the presence of two soft masses, $m_{g}$ and $m_{f}$, and the
discussion necessitates working each case apart. The numerical evaluation
also necessitates the handling of potential divergences coming from soft
light-like loop momenta. Experience with the transverse gluon damping rate $%
\gamma _{t}(p)$ \cite{AABo} indicates that these latter divergences may
ultimately be brought under control, but the infrared ones would most likely
persist. One interesting aspect to mention is that the order $p/m_{f}$ in $%
\gamma _{\pm }(p)$ does not vanish contrary to the gluonic case. Preliminary
results \cite{ABD} tend to indicate that there are no infrared divergences
in the coefficient\footnote{%
We already know from \cite{kobes-kunstatter-mak,braaten-pisarski (quarks)}
that the first coefficient $a_{0}$ is safe.} $a_{1}$ but they tend to appear
in $a_{2}$. On the one hand, this puts `some water' into our arguments
regarding the acceptability of our early expansion in powers of the very
soft external momentum. On the other, recalling that the infrared
divergences we find in the gluonic sector start also at order $\left(
p/m_{g}\right) ^{2}$ \cite{AAB,AA,AABo}, it would be interesting to try to
understand why this is so.

Finally, it is useful to recall that the intensive use of the damping rates
as a mean to probe the properties of finite-$T$ gauge theories is also due
to the fact that in general, calculations in the full HTL-summed
perturbation are quite difficult beyond lowest order. The damping rates are
the simplest such non-trivial quantities to handle. Indeed, though they come
from one-loop graphs with dressed propagators and vertices, they are defined
through the imaginary part of the effective self-energies, something that
simplifies significantly the calculation. The only attempt to correct in
HTL-summed perturbation the quasiparticle spectra to order $g^{2}T$ we are
aware of is that of \cite{shultz}.

This paper is organized as follows. In the next section, we recall the
definition of the quark damping rates and write them in the context of the
HTL-summed perturbation. Their determination amounts to that of the
imaginary part of the next-to-leading order quark self-energy which we carry
in detail in section four. Section three is devoted to discussing the
expansion of the effective self-energy in powers of the external momentum
and to the manner with which we regularize the infrared region. A comparison
with other computational and regularization schemes is also carried, most
particularly those shifting the pole of the effective gluon propagators with
a magnetic mass. The final results are presented and discussed in the last
section.

\section{Quark damping rates in HTL-summed perturbation}

We use the imaginary-time formalism in which the euclidean momentum of the
quark is $P^{\mu }=(p_{0},{\bf p})$ such that $P^{2}=p_{0}^{2}+p^{2}$ with
the fermionic Matsubara frequency$\ p_{0}=(2n+1)\pi T$, $n$ an integer.
Real-time amplitudes are obtained via the analytic continuation $%
p_{0}=-i\omega +0^{+}$ where $\omega $ is the energy of the quark. A\
momentum is said to be soft if both $\omega $ and $p$ are of order $gT$; it
is said to be hard if one is or both are of order $T$. The {\it three}%
-momentum ${\bf p}$ of the on-shell quark is said to be {\it very} soft if $%
p $ is much smaller than $gT$, say of the order $g^{2}T$ and smaller. We
follow closely the notation of \cite{BPhtl} and the HTL results we quote in
this section can all be found there, see also \cite{frenkel-taylor,le bellac}%
.

The effective propagator for the quark can be written as:

\begin{equation}
^{\ast }\Delta _{F}\left( P\right) =-\left[ {\bf \gamma }_{+p}\Delta
_{+}\left( P\right) +{\bf \gamma }_{-p}\Delta _{-}\left( P\right) \right] ,
\label{delta-f}
\end{equation}
where $\gamma ^{\mu }$ are the euclidean Dirac matrices, ${\bf \gamma }_{\pm
p}=\left( \gamma ^{0}\pm i{\bf \gamma }.\widehat{{\bf p}}\right) /2$ and $%
\Delta _{\pm }=\left( D_{0}\mp D_{s}\right) ^{-1}$ with: 
\begin{eqnarray}
D_{0} &=&ip_{0}-\frac{m_{f}^{2}}{p}Q_{0}\left( \frac{ip_{0}}{p}\right) ; 
\nonumber \\
D_{s} &=&p+\frac{m_{f}^{2}}{p}\left[ 1-\frac{ip_{0}}{p}Q_{0}\left( \frac{%
ip_{0}}{p}\right) \right] ,  \label{D0-Ds}
\end{eqnarray}
where the quark thermal mass is $m_{f}=\sqrt{C_{f}/8}\,gT$ and $Q_{0}\left(
x\right) =\frac{1}{2}\ln 
{\displaystyle{x+1 \over x-1}}%
$. The poles of $\Delta _{\pm }(-i\omega ,{\bf p})$ determine the dispersion
laws\footnote{$\left( +\right) $ for real quarks and $\left( -\right) $ for
`plasminos' \cite{braaten}, only thermally excited quasiparticles.} $\omega
_{\pm }(p)$ to lowest order in $g$. For soft quarks, one has: 
\begin{eqnarray}
\omega _{\pm }(p) &=&m_{f}\left[ 1\pm \frac{p}{3m_{f}}+\frac{1}{3}\left( 
\frac{p}{m_{f}}\right) ^{2}\mp \frac{16}{135}\left( \frac{p}{m_{f}}\right)
^{3}+\frac{1}{54}\left( \frac{p}{m_{f}}\right) ^{4}\right.  \nonumber \\
&&\qquad \left. \pm \frac{32}{2835}\left( \frac{p}{m_{f}}\right) ^{5}-\frac{%
139}{12150}\left( \frac{p}{m_{f}}\right) ^{6}\pm \dots \right] .
\label{omega-pm}
\end{eqnarray}

At this lowest order $gT$, $\omega _{\pm }(p)$ are real and the quarks are
not damped. To get the damping rates to their lowest order, one has to
include in the dispersion relations the contribution from the effective
quark self-energy $^{\ast }\Sigma (P)$. Therefore, the inverse of the quark
propagator becomes:

\begin{equation}
\Delta _{F}^{-1}\left( P\right) =\,^{\ast }\Delta _{F}^{-1}\left( P\right)
-\,^{\ast }\Sigma (P)\,.  \label{full-propa-a}
\end{equation}
The effective quark self-energy has also the decomposition $^{\ast }\Sigma
=\gamma ^{0}\,^{\ast }D_{0}+i{\bf \gamma }.\widehat{{\bf p}}\,^{\ast }D_{s}$
where $^{\ast }D_{0}$ and $^{\ast }D_{s}$ are the two functions to be
determined in HTL-summed perturbation. The inverse of the quark propagator
is then: 
\begin{equation}
\Delta _{F}^{-1}\left( P\right) =-\left[ \gamma ^{0}\left( D_{0}+\,^{\ast
}D_{0}\right) +i{\bf \gamma }.\widehat{{\bf p}}\left( D_{s}+\,^{\ast
}D_{s}\right) \right] .  \label{full-propa-b}
\end{equation}

The damping rates for quarks are $\gamma _{\pm }(p)\equiv -{\rm Im}\Omega
_{\pm }\left( p\right) $ where $\Omega _{\pm }$ are the poles of $\Delta
_{F}(-i\Omega ,{\bf p})$. Since the self-energy $^{\ast }\Sigma $ is $g$%
-times smaller than $^{\ast }\Delta _{F}^{-1},$ we have to lowest order: 
\begin{equation}
\gamma _{\pm }\left( p\right) =\left. \frac{{\rm Im\,}^{\ast }f_{\pm }\left(
-i\omega ,p\right) }{\partial _{\omega }f_{\pm }\left( -i\omega ,p\right) }%
\right| _{\omega =\omega _{\pm }(p)+i0^{+}},  \label{gamma-pm-a}
\end{equation}
where $f_{\pm }=D_{0}\mp D_{s}$, $^{\ast }f_{\pm }=\,^{\ast }D_{0}\mp
\,^{\ast }D_{s}$ and $\partial _{\omega }$ stands for $\partial /\partial
\omega $. Using the expressions in (\ref{D0-Ds}), it is easy to expand the
denominator in the above relation in powers of $p/m_{f}$. One then obtains:

\begin{equation}
\gamma _{\pm }\left( p\right) =\left. \frac{1}{2}\left[ 1\pm \frac{2}{3}%
\frac{p}{m_{f}}-\frac{2}{9}\left( \frac{p}{m_{f}}\right) ^{2}+\dots \right]
\,{\rm Im\,}^{\ast }f_{\pm }\left( -i\omega ,p\right) \right| _{\omega
=\omega _{\pm }+i0^{+}}\,.  \label{gamma-pm-b}
\end{equation}
We see that determining $\gamma _{\pm }\left( p\right) $ to lowest order in $%
g$ amounts to calculating the imaginary part of the next-to-leading order
quark self-energy.

The HTL-summed perturbation \cite{BPhtl,frenkel-taylor,le bellac} dictates
that the next-to-leading order quark self-energy is given in imaginary-time
formalism by: 
\begin{equation}
^{\ast }\Sigma \left( P\right) =\,^{\ast }\Sigma _{1}\left( P\right)
+\,^{\ast }\Sigma _{2}\left( P\right) ,  \label{eff-self-energy}
\end{equation}
where we have: 
\begin{equation}
^{\ast }\Sigma _{1}\left( P\right) =-g^{2}C_{f}{\rm Tr}_{{\rm soft}}\left[
^{\ast }\Gamma ^{\mu }\left( P,-Q;-K\right) \,^{\ast }\Delta _{F}\left(
Q\right) \,^{\ast }\Gamma ^{\nu }\left( -P,Q;K\right) \,^{\ast }\Delta _{\mu
\nu }\left( K\right) \right] ,  \label{eff-self-energy1}
\end{equation}
and: 
\begin{equation}
^{\ast }\Sigma _{2}\left( P\right) =-\frac{i}{2}g^{2}C_{f}{\rm Tr}_{{\rm soft%
}}\left[ ^{\ast }\widetilde{\Gamma }^{\mu \nu }\left( P,-P;K,-K\right)
\,^{\ast }\Delta _{\mu \nu }\left( K\right) \right] .
\label{eff-self-energy2}
\end{equation}
$K$ is the soft gluon loop momentum, $Q=P-K$ and ${\rm Tr}\equiv T%
\mathrel{\mathop{\sum }\limits_{k_{0}}}%
\int 
{\displaystyle{d^{3}k \over (2\pi )^{3}}}%
$ with $k_{0}=2n\pi T$, a bosonic Matsubara frequency. The subscript
``soft'' means that only soft values of $K$ are allowed in the integrals;
hard values have dressed the propagators and vertices. Note that since the
loop momentum $K$ is soft, both propagators and vertices involved in (\ref
{eff-self-energy}) must be dressed.

The effective gluonic propagator $^{\ast }\Delta _{\mu \nu }\left( K\right) $
is taken in the strict Coulomb gauge where it has a simplified structure. It
is given by $^{\ast }\Delta _{00}\left( K\right) =\,^{\ast }\Delta
_{l}\left( K\right) $, $^{\ast }\Delta _{0i}\left( K\right) =0$ and $^{\ast
}\Delta _{ij}\left( K\right) =\left( \delta _{ij}-\widehat{k}_{i}\widehat{k}%
_{j}\right) \,^{\ast }\Delta _{t}\left( K\right) $ with $^{\ast }\Delta _{l}$
and $^{\ast }\Delta _{t}$ having the following expressions: 
\begin{equation}
^{\ast }\Delta _{l}\left( K\right) =\frac{1}{k^{2}-\delta \Pi _{l}\left(
K\right) }\,;\,\,\,\,\,\,\,\,\,\,\,\,\,\,\,\,\,\,\,\,^{\ast }\Delta
_{t}\left( K\right) =\frac{1}{K^{2}-\delta \Pi _{t}\left( K\right) }\,,
\label{propa-glu}
\end{equation}
where $\delta \Pi _{l}(K)=3m_{g}^{2}Q_{1}(\frac{ik_{0}}{k})$ and $\delta \Pi
_{t}(K)=\frac{3}{5}m_{g}^{2}\left[ Q_{3}(\frac{ik_{0}}{k})-Q_{1}(\frac{ik_{0}%
}{k})-\frac{5}{3}\right] $. $Q_{i}(\frac{ik_{0}}{k})$ is a Legendre function
of the second kind and the gluon thermal mass $m_{g}$=$\sqrt{N_{c}+N_{f}/2}%
\,gT/3$. The effective (dressed) vertices $^{\ast }\Gamma $ intervening in (%
\ref{eff-self-energy}) are of the form: 
\begin{equation}
^{\ast }\Gamma =\Gamma +\delta \Gamma \,,  \label{effe-verti}
\end{equation}
where $\Gamma $ is the bare (tree) vertex and $\delta \Gamma $ is the
corresponding hard thermal loop. The two effective vertices that enter the
calculation of the effective self-energy (\ref{eff-self-energy}) are the
effective quark-gluon vertex: 
\begin{equation}
^{\ast }\Gamma ^{\mu }(P,Q;R)=\gamma ^{\mu }+m_{f}^{2}\int \frac{d\Omega _{s}%
}{4\pi }\frac{S^{\mu }S\hspace{-0.6183pc}/}{PS\,QS}\,,  \label{eff-ver-1}
\end{equation}
where the second term is the hard thermal loop, and the effective
two-gluons-quark-antiquark vertex: 
\begin{equation}
^{\ast }\widetilde{\Gamma }^{\mu \nu }(P,-P;K,-K)=-2m_{f}^{2}\int \frac{%
d\Omega _{s}}{4\pi }\frac{S^{\mu }S^{\nu }S\hspace{-0.6183pc}/}{PS\left(
P+K\right) \hspace{-2pt}S\left( P-K\right) \hspace{-2pt}S}\,.
\label{eff-ver-2}
\end{equation}
Note that the bare two-gluons-quark-antiquark vertex is zero so that the
corresponding effective vertex is just the hard thermal loop. In both (\ref
{eff-ver-1}) and (\ref{eff-ver-2}), $S\equiv (i,\widehat{{\bf s}})$ and $%
\Omega _{s}$ is the solid angle of $\widehat{{\bf s}}$.

The task is to attempt to get an expression for the imaginary part of the
effective quark self-energy $^{\ast }\Sigma \left( P\right) $. The `natural'
sequence of steps to follow is first to perform the angular integrations in
the dressed vertices (\ref{eff-ver-1}) and (\ref{eff-ver-2}). Next is to do
the Matsubara sum in (\ref{eff-self-energy}). Only then the continuation to
real quark energies $p_{0}=-i\omega +0^{+}$ can be taken and the on-shell
condition enforced. Last is to find a way to perform the integration over
the gluon loop three-momentum ${\bf k}$. However, given the complicated
expressions we are faced with, it is practically very difficult to follow
this sequence of operations. What we do in this work is first expand the
effective self-energy in powers of the quark momentum $p/m_{f}$. This allows
for an easy angular integration over $\Omega _{s}$. Only then do we perform
the Matsubara sum, this by using the spectral representation of the
different quantities involved. The angular integration over $\Omega _{k}$ is
subsequently done and the remaining integrals can be calculated numerically.
We discuss this procedure in more detail in the next section.

\section{Regularization and expansion in external momentum}

The expansion in powers of the external momentum of the HTL-summed
next-to-leading order self-energies can be questioned from the outset in
view of the fact that infrared divergences do appear in next-to-leading
order physical quantities like the damping rates \cite{AA}. Are these
divergences genuine or merely artifacts due to the method used? First,
recall that we are considering only very soft external momenta. Note also
that the HTL\ framework itself allows for an expansion of quantities in
powers of the soft external momenta. For example, the gluon and quark
on-shell energies $\omega (p)$ are obtained in the literature in the form of
a series in powers of soft $p$ \cite{rho,le bellac}. The same is true for
the residue and cut functions intervening in the spectral decomposition of
the effective propagators \cite{AA}. It is therefore legitimate to expect
the perturbation {\it built on} hard thermal loops (these being considered
as a zeroth order approximation) to be analytic in very soft $p,$ and hence
admit an expansion in powers of such momenta. Also, such an expansion is not
proper to this work, it has been previously used in the literature, for
example in \cite{braaten-nieto2}.

There is of course a distinction between the analyticity in $p$ and that in $%
g.$ For example, the standard loop expansion of QCD is in powers of $g^{2}$
whereas in HTL-summed perturbation, the expansion is in powers of $\sqrt{%
g^{2}}$. The same is true for other theories like the prototype $\lambda
\phi ^{4}$ theory \cite{lambda-phi-4} and QED \cite{on QED}. This may
introduce a non-analyticity with respect to $g$ in some quantities, but does
not necessarily change drastically the analytic behavior of these quantities
with respect to very soft $p$. Take for example the estimation of the soft
gluon damping rates made in \cite{pisarski2}: 
\begin{equation}
\gamma _{t,l}(p)\sim -\frac{g^{2}N_{c}T}{4\pi }\ln g\,v_{t,l}(p),
\label{gam-tl-pisarski}
\end{equation}
where $v_{t,l}(p)$ are the corresponding group velocities. These rates are
clearly non-analytic in small $g$, but perfectly analytic in $p$: they even
tend to zero\footnote{%
A non-acceptable limit as we will discuss shortly.} as $p\rightarrow $0.

Regarding the damping rates, our starting position is that the quark-gluon
plasma is to be a stable phase of hadronic matter, at least for very soft
excitations \cite{stability-qgp}. QCD at high temperature in the
(lowest-order) HTL approximation is `finite'. At next-to-leading order,
HTL-summed perturbation yields finite and positive damping rates for
zero-momentum on-shell quarks and transverse gluons. The stability criterion
ensures that we must expect the damping rates to remain finite and positive
for non-zero very soft momenta. This translates into expecting the damping
rates to admit a series expansion in powers of these very soft external
momenta. This of course does not rule out a possible loss of analyticity for
larger values of $p$, even just soft values. That would simply indicate new
physics to explore. But because of the stability criterion, the analyticity
must be preserved for very soft momenta. This is one important check to use
in order to discuss the consistency and completeness of a given
calculational scheme like the HTL-summed perturbation.

Expecting an infrared problem, we introduce an infrared cut-off $\eta >0$
such that $\int_{0}^{+\infty }dk$ in (\ref{eff-self-energy1})\ and (\ref
{eff-self-energy2}) is replaced by $\int_{\eta }^{+\infty }dk$. The cut-off $%
\eta $ is fixed for the rest of the calculation. It is physically useful to
see it as representing the magnetic scale $g^{2}T$. This means that $k$ is
never smaller than $\eta $. In other words, we are summing contributions
from all soft momenta\footnote{%
Hard momenta are already summed in the hard thermal loops.} $k$ but not the
very soft ones, i.e., those smaller than $\eta $. We {\it always} regard the
external momentum $p$ as smaller than $\eta $. We are therefore always
working in the kinematic region $0\leq p<\eta \leq k$. This allows for the
expansion in powers of $p$ of all quantities that are functions of $q=\left| 
{\bf p}-{\bf k}\right| $. This is true in particular for $1/QS$ and the
effective propagators $^{\ast }\Delta \left( Q\right) $. The expansion of $%
1/PS$ does not pose a problem in itself.

It is useful to emphasize once more that our calculation sums the
contributions from {\it only} the soft integration momenta $\eta \leq k$:
the very soft momenta $0\leq k<\eta $ are systematically excluded and the
hard region is cut by the spectral densities \cite{AA}. If the integration
is not sensitive to the very soft region (the magnetic sector), then the
subsequent limit $\eta \rightarrow 0$ in the final result should be smooth.
If on the contrary there is sensitivity, it would mean that important
contributions from this sector may be `missed' by the HTL-summed
perturbation. This is the essence of our point. Indeed, recall that
regarding the self-energies, the HTL scheme discusses the two scales $T$
(hard) and $gT$ (soft), whereas with $g$ and $T$, one has a hierarchy of
scales $g^{n}T$ with $n$ a nonnegative integer. As a matter of fact, \cite
{aurenche-gelis-kobes-petitgirard} argues that there may be scales between $%
T $ and $gT$ that play a significant role, and so $n$ may not even be an
integer. Since the HTL scheme, when built, does not consider effects like
magnetic screening which (are believed to) arise nonperturbatively at order $%
g^{2}T$ and are not present at the HTL level, a perturbation built on the
HTL summation may not be able to reproduce them. What it can do is to bring
what contributes from the soft region to the very soft one. Therefore,
excluding the very soft region from the $k$-integration as we do may not be
all unreasonable a thing to do. We think that the presence of magnetic
sensitivity which manifests itself in infrared divergences indicates the
very presence of these magnetic effects that the HTL\ scheme seems to be not
able to accommodate.

Let us compare our calculation of the damping rates with the estimation (\ref
{gam-tl-pisarski}) mentioned above. This latter is different in many
respects from the one we carry. It is obtained in the kinematic region where
the loop momentum $k$ is restricted to the very soft region ($0\leq k<\eta $
in our notation) whereas $p$ is just soft, of order $gT$ \cite{pisarski2}.
In some sense, there, it is the very soft momenta that are integrated out;
the soft ones are disregarded, something opposite to what we do. Result (\ref
{gam-tl-pisarski}) cannot be carried to the very soft region $p<\eta $, in
particular to the point $p\rightarrow 0$ for it will give zero (using the
expressions of the group velocities at very soft momenta) whereas the
damping rates there are finite. At the same time, our results can never be
carried to the region $p>\eta $. Clearly then, it should not be problematic
if different analytic results are obtained. In fact, if really different
results are obtained, which is the case, it only constitutes a further
indication of the sensitivity of the HTL\ scheme to the magnetic sector.
Equally interesting to note in the estimation (\ref{gam-tl-pisarski}) is
that, in order to screen the divergent behavior at very soft momenta $k$, a
regularization is used, which amounts to introducing a magnetic mass $m_{%
{\rm mag}}$ in the otherwise divergent propagators. It is clear that
screening of chromomagnetic fields, if it occurs, is not necessarily going
to manifest itself by a simple shift of the pole in the corresponding
propagator by a momentum independent magnetic mass \cite{rebhan2}.

The presence of magnetic effects not handled by the HTL scheme is discussed
in \cite{braaten-nieto2}. It is argued there that for distances to order $%
1/T $, ordinary perturbation (the standard loop expansion) is reliable. For
distances to order $1/gT$, the effective theory that screens static
chromoelectric fields (the Braaten-Pisarski scheme) is reliable and it can
be treated in perturbation. However, for distances to order $1/g^{2}T$, one
needs another effective theory which cannot be treated by perturbation
(treated by lattice simulations for example). This last statement is
emphasized by comparing the asymptotic behavior of the Polyakov-loop
correlator determined from a magnetic lagrangian with that determined from
an electric lagrangian with a put-by-hand magnetic mass $m_{{\rm mag}}$.
With the magnetic lagrangian, an exponential decay governed by the lowest
glueball state is obtained, an asymptotic behavior different from the one
obtained from the magnetic-mass enhanced electric lagrangian. A comparison
of these results with lattice simulations indicates that the glueball-state
result is more compatible with the lattice ones. One interesting inference
one can draw from the above comments is that regularizing the infrared
sector in HTL perturbation with a simple magnetic mass may not be the best
description of magnetic effects, in particular if those are not incorporated
in the scheme itself.

It is important to stress that from a pure computational standpoint, matters
are not straightforward if we defer in the self-energies the expansion in $p$
after the angular integrals. Indeed, one has to deal with expressions quite
complicated and involved, something of the sort $T\sum_{k_{0}}\int
d^{3}k\int d\Omega _{s_{1}}\int d\Omega _{s_{2}}\frac{f(\widehat{{\bf s}}%
_{1},\widehat{{\bf s}}_{2})}{PS_{1}\,KS_{1}\,PS_{2}\,QS_{2}}\,^{\ast }\Delta
\left( K\right) \,^{\ast }\Delta \left( Q\right) $. The Matsubara sum can be
performed if the effective propagators, $\frac{1}{KS_{1}}$ and $\frac{1}{%
QS_{2}}$ are replaced by their respective spectral decomposition, see below.
But this will bring in more than one energy denominator, which would
compromise the straightforward extraction of the imaginary part of the
effective self-energy. More serious a problem is the subsequent angular
integration which will be very difficult, it not impossible, to perform \cite
{frenkel-taylor}.

Finally, it turns out that for quarks, the expansion is in powers of $%
p/m_{f} $ and not in $\left( p/m_{f}\right) ^{2}$, as is the case for gluons
(where $m_{f}$ is replaced by $m_{g}$), \cite{AAT,AAB,AA}. Preliminary
results \cite{ABD} tend to indicate that the second coefficient (that of $%
p/m_{f}$) is infrared safe together with the first one. This may suggest
then that the expansion in powers of $p$ is not sole to `blame' for
obtaining infrared divergent damping rates; other effects may be in play.

\section{Imaginary part of one-loop HTL-summed quark self-energy}

Now we present the calculation of the HTL-summed next-to-leading order quark
self-energy from which we extract the imaginary part. We first describe how
we get an expression for $%
\mathop{\rm Im}%
\,^{\ast }\Sigma _{1}\left( P\right) $ defined in (\ref{eff-self-energy1})
and then for $%
\mathop{\rm Im}%
\,^{\ast }\Sigma _{2}\left( P\right) $ defined in (\ref{eff-self-energy2}).
From now on, we take $m_{f}=1$. This will simplify the final expressions we
obtain. There remains another soft mass in the problem, $m_{g}$, and so we
define $m=m_{g}/m_{f}=\frac{4}{3}\sqrt{\frac{N_{c}(N_{c}+N_{f}/2)}{%
N_{c}^{2}-1}}$. It is easy to see that we always have $m>1$.

\subsection{Calculation of $%
\mathop{\rm Im}%
\,^{\ast }\Sigma _{1}\left( P\right) $}

Using the structure of the fermion propagator (\ref{delta-f}) and that of
the gluon propagator in the strict Coulomb gauge given just before (\ref
{propa-glu}), we see that $^{\ast }\Sigma _{1}\left( P\right) $ is composed
of four terms: 
\begin{eqnarray}
^{\ast }\Sigma _{1}(P) &=&\frac{8}{T^{2}}\sum_{\varepsilon =\pm
}\,T\sum_{k_{0}}\int \frac{d^{3}k}{\left( 2\pi \right) ^{3}}\left[ ^{\ast
}\Gamma ^{0}\left( P,-Q;-K\right) \Delta _{\varepsilon }\left( Q\right) {\bf %
\gamma }_{\varepsilon q}\,^{\ast }\Gamma ^{0}\left( -P,Q;K\right) \,^{\ast
}\Delta _{l}\left( K\right) \right.   \nonumber \\
&&+\,^{\ast }\Gamma ^{i}\left( P,-Q;-K\right) \Delta _{\varepsilon }\left(
Q\right) {\bf \gamma }_{\varepsilon q}\,^{\ast }\Gamma ^{j}\left(
-P,Q;K\right) \,(\delta _{ij}-\widehat{k}_{i}\widehat{k}_{j})\,^{\ast
}\Delta _{t}\left( K\right) \,.  \label{sigma-star-1}
\end{eqnarray}
The first two terms denoted $^{\ast }\Sigma _{\varepsilon l}(P)$, those with
the longitudinal gluon propagator, are calculated separately from the two
others denoted $^{\ast }\Sigma _{\varepsilon t}(P)$. We will illustrate the
different steps of the calculation for $^{\ast }\Sigma _{-l}(P)$. Using the
definition of the effective vertex (\ref{eff-ver-1}) and making the change
of integration variable $K\rightarrow P-K$, we have: 
\begin{eqnarray}
^{\ast }\Sigma _{-l}(P) &=&\frac{8}{T^{2}}\,T\widetilde{\sum_{k_{0}}}\int 
\frac{d^{3}k}{\left( 2\pi \right) ^{3}}\left[ {\bf \gamma }_{+k}-\int \frac{%
d\Omega _{s}}{4\pi }\,\frac{2iS\hspace{-0.6252pc}/+\gamma ^{0}{\bf \gamma }.%
\widehat{{\bf k}}\,S\hspace{-0.6252pc}/+S\hspace{-0.6252pc}/\,{\bf \gamma }.%
\widehat{{\bf k}}\gamma ^{0}}{2PS\,KS}\right.   \nonumber \\
&&\left. -\int \frac{d\Omega _{s_{1}}}{4\pi }\int \frac{d\Omega _{s_{2}}}{%
4\pi }\frac{S\hspace{-0.6252pc}/_{1}\,{\bf \gamma }_{-k}\,S\hspace{-0.6252pc}%
/_{2}}{PS_{1}\,KS_{1}\,PS_{2}\,KS_{2}}\right] \,\Delta _{-}\left( K\right)
\,^{\ast }\Delta _{l}\left( Q\right) \,.  \label{sigma-l-1}
\end{eqnarray}
The tilde over the sum sign indicates that $k_{0}$ is fermionic.

Let us start with $I_{1}$, the term in (\ref{sigma-l-1}) where there is one
solid-angle integral over $\Omega _{s}$. This latter is carried in a
reference frame where $\widehat{{\bf k}}$ is the principle axis (i.e., the `$%
z$-axis'). The solid angle is then $\Omega _{s}=(\theta ,\varphi )$ such
that $\widehat{{\bf k}}.\widehat{{\bf s}}=\cos \theta $ and $\widehat{{\bf p}%
}.\widehat{{\bf s}}=\cos \psi \cos \theta -\sin \psi \sin \theta \sin
\varphi $, where $\cos \psi =\widehat{{\bf k}}.\widehat{{\bf p}}\,$. Also,
we have ${\bf \gamma }.\widehat{{\bf s}}=\gamma ^{\prime 1}\sin \theta \cos
\varphi +\gamma ^{\prime 2}(\sin \psi \cos \theta +\cos \psi \sin \theta
\sin \varphi )+\gamma ^{\prime 3}(\cos \psi \cos \theta -\sin \psi \sin
\theta \sin \varphi )$, where $\left\{ \gamma ^{\prime i}\right\} $ are the
three spatially rotated Dirac matrices written in a reference frame where $%
\widehat{{\bf p}}$ is the principle axis and $\widehat{{\bf k}}$ in the $%
(y,z)$-plane. They are fixed in the integration over $\Omega _{s}$.
Performing all the (anti)commutations, we have: 
\begin{equation}
I_{1}=\frac{8}{T^{2}}\,T\widetilde{\sum_{k_{0}}}\int \frac{d^{3}k}{\left(
2\pi \right) ^{3}}\int \frac{d\Omega _{s}}{4\pi }\frac{\gamma ^{0}(1-\cos
\theta )+i{\bf \gamma }.\widehat{{\bf k}}-i{\bf \gamma }.\widehat{{\bf s}}}{%
PS\,KS}\Delta _{-}\left( K\right) \,^{\ast }\Delta _{l}\left( Q\right) \,.
\label{one-solid-angle-in-sigma-plus-l}
\end{equation}
In order to be able to perform with ease the above solid-angle integral, we
use the expansion: 
\begin{equation}
{\displaystyle{1 \over PS}}%
=%
{\displaystyle{1 \over ip_{0}}}%
\left[ 1-%
{\displaystyle{{\bf p}.\widehat{{\bf s}} \over ip_{0}}}%
-%
{\displaystyle{{\bf p}.\widehat{{\bf s}}^{2} \over p_{0}^{2}}}%
+\dots\right] \,.  \label{expan-PS}
\end{equation}
This expansion is valid in the region $p<\left| ip_{0}\right| $, a condition
always satisfied before analytic continuation and after. Before because $%
p_{0}=\left( 2n+1\right) \pi T$ and $p\sim g^{2}T$. After because for very
soft momenta, $ip_{0}=m_{f}+O(p/m_{f})\sim gT$, see (\ref{omega-pm}).\ The
solid-angle integral in (\ref{one-solid-angle-in-sigma-plus-l}) then reads: 
\begin{eqnarray*}
&&\frac{1}{ip_{0}}\int_{0}^{2\pi }d\varphi \int_{0}^{\pi }d\theta \,\sin
\theta \frac{\gamma ^{0}(1-\cos \theta )+i{\bf \gamma }.\widehat{{\bf k}}-i%
{\bf \gamma }.\widehat{{\bf s}}}{ik_{0}+k\cos \theta } \\
&&\times \left[ 1-\frac{p}{ip_{0}}(\cos \psi \cos \theta -\sin \psi \sin
\theta \sin \varphi )-\frac{p^{2}}{p_{0}^{2}}(\cos \psi \cos \theta -\sin
\psi \sin \theta \sin \varphi )^{2}+\dots \right] .
\end{eqnarray*}
The angular integrations are now straightforward and we obtain: 
\begin{eqnarray}
I_{1} &=&\frac{8}{T^{2}}\,T\widetilde{\sum_{k_{0}}}\int \frac{d^{3}k}{\left(
2\pi \right) ^{3}}\frac{1}{ip_{0}k}\left[ \gamma ^{0}\left[ -1+\left( 1+%
\frac{ik_{0}}{k}\right) Q_{0k}-\frac{px}{ip_{0}}\left( 1+\frac{ik_{0}}{k}%
\right) \left( 1-\frac{ik_{0}}{k}Q_{0k}\right) \right. \right.  \nonumber \\
&&\hspace{-0.35in}+\frac{p^{2}}{p_{0}^{2}}\left[ x^{2}\left( \frac{1}{3}+%
\frac{ik_{0}}{k}-\frac{k_{0}^{2}}{k^{2}}+\frac{k_{0}^{2}}{k^{2}}\left( 1+%
\frac{ik_{0}}{k}\right) Q_{0k}\right) -\frac{1}{2}\left( 1-x^{2}\right)
\left( -\frac{2}{3}+\frac{ik_{0}}{k}-\frac{k_{0}^{2}}{k^{2}}\right. \right. 
\nonumber \\
&&\hspace{-0.35in}+\hspace{-0.1in}\left. \left. \left. \left( 1+\frac{ik_{0}%
}{k}\right) \left( 1+\frac{k_{0}^{2}}{k^{2}}\right) Q_{0k}\right) \right] %
\right] +i{\bf \gamma }.\widehat{{\bf k}}\left[ Q_{0k}-\frac{px}{ip_{0}}%
\left( 1-\frac{ik_{0}}{k}Q_{0k}\right) \right. +\frac{p^{2}}{p_{0}^{2}}\left[
x^{2}\frac{ik_{0}}{k}\left( 1-\frac{ik_{0}}{k}Q_{0k}\right) \right. 
\nonumber \\
&&\hspace{-0.35in}-\left. \frac{1}{2}(1-x^{2})\left( \frac{ik_{0}}{k}+\left(
1+\frac{k_{0}^{2}}{k^{2}}\right) Q_{0k}\right) \right] -i\gamma ^{\prime
2}\sin \psi \left[ 1-\frac{ik_{0}}{k}+\frac{px}{2ip_{0}}\left( 3\frac{ik_{0}%
}{k}+\left( 1+3\frac{k_{0}^{2}}{k^{2}}\right) Q_{0k}\right) \right. 
\nonumber \\
&&\hspace{-0.35in}+\frac{p^{2}}{p_{0}^{2}}\left[ x^{2}\left( \frac{1}{3}+2%
\frac{k_{0}^{2}}{k^{2}}-\frac{ik_{0}}{k}\left( 1+2\frac{k_{0}^{2}}{k^{2}}%
\right) Q_{0k}\right) \right. -\left. \left. \frac{1}{2}(1-x^{2})\left( 
\frac{2}{3}+\frac{k_{0}^{2}}{k^{2}}-\frac{ik_{0}}{k}\left( 1+\frac{k_{0}^{2}%
}{k^{2}}\right) Q_{0k}\right) \right] \right]  \nonumber \\
&&\hspace{-0.35in}-i\gamma ^{\prime 3}\left[ x\left( 1-\frac{ik_{0}}{k}%
Q_{0k}\right) \right. +\frac{p}{ip_{0}}\left[ x^{2}\left( \frac{ik_{0}}{k}+%
\frac{k_{0}^{2}}{k^{2}}Q_{0k}\right) -\frac{1}{2}(1-x^{2})\left( \frac{ik_{0}%
}{k}+\left( 1+\frac{k_{0}^{2}}{k^{2}}\right) Q_{0k}\right) \right]  \nonumber
\\
&&\hspace{-0.35in}\left. +\left. \frac{p^{2}}{p_{0}^{2}}\left[ x^{3}\left( 
\frac{ik_{0}}{3k}-\frac{ik_{0}^{3}}{k^{3}}-\frac{k_{0}^{4}}{k^{4}}%
Q_{0k}\right) -\frac{3}{2}x(1-x^{2})\left( \frac{2}{3}+\frac{k_{0}^{2}}{k^{2}%
}-\frac{ik_{0}}{k}\left( 1+\frac{k_{0}^{2}}{k^{2}}\right) Q_{0k}\right) %
\right] \right] +\dots \right]  \nonumber \\
&&\hspace{-0.35in}\times\Delta _{-}\left( K\right) \,^{\ast }\Delta
_{l}\left( Q\right) \,,
\label{one-solid-angle-in-sigma-plus-l--intermediary1}
\end{eqnarray}
where $x=\cos \psi $ and $Q_{0k}$ stands for $Q_{0}(ik_{0}/k).$

The next step is to perform the integrals over the solid angle of $\widehat{%
{\bf k}}$ in a reference frame where $\widehat{{\bf p}}$ is the principle
axis. For this, it is most useful to develop all functions of $q=\left| {\bf %
p}-{\bf k}\right| $ around $k$ for (very) small $p$. The validity of these
expansions is discussed in the previous section. In particular, here we
need: 
\begin{equation}
^{\ast }\Delta _{l,t}\left( q_{0},q\right) =\left[ 1-px\,\partial _{k}+\frac{%
p^{2}}{2}\left( \frac{1-x^{2}}{k}\partial _{k}+x^{2}\partial _{k}^{2}\right)
+\dots \right] \,^{\ast }\Delta _{l,t}\left( q_{0},k\right) \,.
\label{expansion-delta-tl}
\end{equation}
The solid angle of $\widehat{{\bf k}}$ is $\Omega _{k}=(\psi ,\alpha )$ and
we have the relation: 
\begin{equation}
\gamma ^{\prime 1}=\gamma ^{1}\cos \alpha -\gamma ^{2}\sin \alpha ;\quad
\gamma ^{\prime 2}=\gamma ^{1}\sin \alpha +\gamma ^{2}\cos \alpha ;\quad
\gamma ^{\prime 3}=\gamma ^{3}\,,  \label{dirac-matrices}
\end{equation}
where the $\left\{ \gamma ^{i}\right\} $ are the (fixed) spatial Dirac
matrices. Using (\ref{expansion-delta-tl}), the integrations over $\psi $
and $\alpha $ become straightforward. We obtain: 
\begin{eqnarray}
I_{1} &=&\frac{4}{\pi ^{2}T^{2}}\,T\widetilde{\sum_{k_{0}}}\int_{\eta
}^{+\infty }dk\frac{k\,}{ip_{0}}\Delta _{-K}\left[ \gamma ^{0}\left[
-1+\left( 1+\frac{ik_{0}}{k}\right) Q_{0k}+\frac{p^{2}}{3}\left[ \frac{1}{%
p_{0}^{2}}\left( 1-\left( 1+\frac{ik_{0}}{k}\right) Q_{0k}\right) \right.
\right. \right.  \nonumber \\
&&\left. +\left. \frac{1}{ip_{0}}\left( 1+\frac{ik_{0}}{k}\right) \left( 1-%
\frac{ik_{0}}{k}Q_{0k}\right) \partial _{k}-\left( 1-\left( 1+\frac{ik_{0}}{k%
}\right) Q_{0k}\right) \left( \frac{1}{k}\partial _{k}+\frac{1}{2}\partial
_{k}^{2}\right) \right] \right]  \nonumber \\
&&\left. +\frac{1}{3}i{\bf \gamma }^{3}p\left( 1-\left( 1+\frac{ik_{0}}{k}%
\right) Q_{0k}\right) \left( -\frac{1}{ip_{0}}+\partial _{k}\right) +\dots %
\right] \,^{\ast }\Delta _{l}\left( q_{0},k\right) \,,
\label{one-solid-angle-in-sigma-plus-l--intermediary2}
\end{eqnarray}
where $\Delta _{-K}$ stands for $\Delta _{-}(K)$. Note the introduction of
the infrared cut-off $\eta $. Note also that terms proportional to $p$ do
not vanish, contrary to what happens for gluons \cite{AAT,AAB}.

The next step for $I_{1}$ is to perform the Matsubara sum. This will be done
after we get for the first term $I_{0}$ in $^{\ast }\Sigma _{-l}(P)$ (the
one that involves no angular integrals) and the third term $I_{2}$ (the one
that involves two such integrals) expressions similar to (\ref
{one-solid-angle-in-sigma-plus-l--intermediary2}). As for $I_{0}$, the
calculation is simpler: only an integral over $\Omega _{k}$ using the
expansion (\ref{expansion-delta-tl}) is needed. We get: 
\begin{equation}
I_{0}=\frac{2}{3\pi ^{2}T^{2}}\,T\widetilde{\sum_{k_{0}}}\int_{\eta
}^{+\infty }dk\,k^{2}\Delta _{-K}\left[ \gamma ^{0}\left[ 3+p^{2}\left( 
\frac{1}{k}\partial _{k}+\frac{1}{2}\partial _{k}^{2}\right) \right]
-i\gamma ^{3}p\partial _{k}+\dots \right] \,^{\ast }\Delta _{l}\left(
q_{0},k\right) \,.  \label{no-solid-angle-in-sigma-plus-l--intermediary1}
\end{equation}
As for $I_{2}$, more work is needed. But we are fortunate here since $%
\widehat{{\bf s}}_{1}$ is not `coupled' to $\widehat{{\bf s}}_{2}$ so that
each solid-angle integral can be performed independently from the other.
Each integral is thus performed along the lines shown for $I_{1}$, and so,
there is no need to re-display the steps. After the two integrations are
done, we multiply the two results, keeping terms to order $p^{2}$ only and
taking care of the Dirac algebra. We obtain: 
\begin{eqnarray}
I_{2} &=&\frac{2}{\pi ^{2}T^{2}}T\widetilde{\sum_{k_{0}}}\int_{\eta
}^{+\infty }\frac{dk}{p_{0}^{2}}\Delta _{-K}\left[ \hspace{-0.05in}-\gamma
^{0}a_{-}^{2}+i\gamma ^{3}\frac{p}{3}a_{-}^{2}\left( \hspace{-0.05in}-\frac{2%
}{ip_{0}}+\hspace{-0.05in}\partial _{k}\right) +\gamma ^{0}\frac{p^{2}}{3}%
\left[ \hspace{-0.02in}\frac{1}{2p_{0}^{2}}\left( \hspace{-0.02in}3-\left( 
\hspace{-0.05in}2-6\frac{ik_{0}}{k}\right) a_{-}\right. \right. \right. 
\nonumber \\
&&+\left. \left. \left. \left( 5-2\frac{ik_{0}}{k}-3\frac{k_{0}^{2}}{k^{2}}%
\right) a_{-}^{2}\right) +\frac{2}{ip_{0}}a_{-}\left( 1+\frac{ik_{0}}{k}%
a_{-}\right) \partial _{k}-a_{-}\left( \frac{1}{k}\partial _{k}+\frac{1}{2}%
\partial _{k}^{2}\right) \right] +\dots \right] \,  \nonumber \\
&&\times ^{\ast }\Delta _{l}\left( q_{0},k\right) ,
\label{two-solid-angles-in-sigma-plus-l--intermediary}
\end{eqnarray}
where we have denoted for short $a_{\varepsilon }=1+\varepsilon \left(
1-\varepsilon \frac{ik_{0}}{k}\right) Q_{0k}\,$, $\varepsilon =\pm $. We can
now put together $I_{0}$, $I_{1}$ and $I_{2}$ to get a first expression for $%
^{\ast }\Sigma _{-l}(P)$. Since $^{\ast }\Sigma _{+l}(P)$ is calculated in
the same way and the only differences are mere signs, it is more economical
to write the result for both terms in one single expression. We find: 
\begin{eqnarray}
^{\ast }\Sigma _{\varepsilon l}(P) &=&\frac{2}{\pi ^{2}T^{2}}\,T\widetilde{%
\sum_{k_{0}}}\int_{\eta }^{+\infty }dk\,k^{2}\Delta _{\varepsilon K}\left[
\gamma ^{0}\left( 1+\frac{2\varepsilon }{ip_{0}k}a_{\varepsilon }-\frac{1}{%
p_{0}^{2}k^{2}}a_{\varepsilon }^{2}\right) \right.  \nonumber \\
&&\hspace{-0.65in}+i\gamma ^{3}\frac{p}{3}\left[ \varepsilon \partial _{k}+%
\frac{2}{ip_{0}k}a_{\varepsilon }\left( \frac{\varepsilon }{ip_{0}}+\partial
_{k}\right) -\frac{\varepsilon }{p_{0}^{2}k^{2}}a_{\varepsilon }^{2}\left( 
\frac{2\varepsilon }{ip_{0}}+\partial _{k}\right) \right]  \nonumber \\
&&\hspace{-0.65in}+\gamma ^{0}\frac{p^{2}}{3}\left[ \left( \frac{1}{k}%
\partial _{k}+\frac{1}{2}\partial _{k}^{2}\right) +\hspace{-0.02in}\frac{2}{%
ip_{0}k}\left( -\frac{\varepsilon }{p_{0}^{2}}a_{\varepsilon }+\frac{1}{%
ip_{0}}\left( 1-\varepsilon \frac{ik_{0}}{k}a_{\varepsilon }\right) \partial
_{k}+\varepsilon a_{\varepsilon }\left( \frac{1}{k}\partial _{k}+\frac{1}{2}%
\partial _{k}^{2}\right) \right) \right.  \nonumber \\
&&\hspace{-0.65in}+\frac{1}{p_{0}^{2}k^{2}}\hspace{-0.02in}\left( \frac{1}{%
2p_{0}^{2}}\left( 3-2\left( 1+3\varepsilon \frac{ik_{0}}{k}\right)
a_{\varepsilon }+\left( \hspace{-0.02in}5+2\varepsilon \frac{ik_{0}}{k}-3%
\frac{k_{0}^{2}}{k^{2}}\right) a_{\varepsilon }^{2}\right) -\frac{%
2\varepsilon }{ip_{0}}a_{\varepsilon }\left( 1-\varepsilon \frac{ik_{0}}{k}%
a_{\varepsilon }\right) \partial _{k}\right.  \nonumber \\
&&\left. \hspace{-0.65in}\left. -\left. a_{\varepsilon }^{2}\left( \frac{1}{k%
}\partial _{k}+\frac{1}{2}\partial _{k}^{2}\right) \right) \right] \hspace{%
-0.02in}+\dots \right] \,^{\ast }\Delta _{l}\left( q_{0},k\right) \,.
\label{sigma-plus-l-2}
\end{eqnarray}

Now we are ready to perform the Matsubara sum over fermionic $k_{0}$%
\noindent $.$ We will need the spectral decomposition of $\Delta
_{\varepsilon }\left( k_{0},k\right) $, $^{\ast }\Delta _{l}\left(
q_{0},k\right) $ and $Q_{0}(ik_{0}/k)$. They are worked out in \cite{rho,le
bellac} and are given by: 
\begin{eqnarray}
\Delta _{\varepsilon }(k_{0},k) &=&\int_{0}^{1/T}d\tau \,e^{ik_{0}\tau
}\int_{-\infty }^{+\infty }d\omega \,\rho _{\varepsilon }(\omega ,k)\left( 1-%
\tilde{n}(\omega )\right) e^{-\omega \tau }\,;  \nonumber \\
\Delta _{t,l}(k_{0},k) &=&\int_{0}^{1/T}d\tau \,e^{ik_{0}\tau }\int_{-\infty
}^{+\infty }d\omega \,\rho _{t,l}(\omega ,k)\left( 1+n(\omega )\right)
e^{-\omega \tau }\,;  \nonumber \\
Q_{0}(ik_{0}/k) &=&\int_{0}^{1/T}d\tau \,e^{ik_{0}\tau }\int_{-\infty
}^{+\infty }d\omega \,\rho _{0}(\omega ,k)\left( 1-\tilde{n}(\omega )\right)
e^{-\omega \tau }\,.  \label{spectral-relations}
\end{eqnarray}
$n(\omega )$ ($\tilde{n}(\omega )$) is the Bose-Einstein (Fermi-Dirac)
distribution and the rho's are the spectral densities. Before replacing
these above quantities, it is first necessary to rearrange terms in (\ref
{sigma-plus-l-2}) in a such a way that products of only two such functions
appear. The reason behind is to ensure the appearance of only one energy
denominator just before the extraction of the imaginary part, see below. For
this purpose, we use the following easy-to-check relations: 
\begin{eqnarray}
a_{\varepsilon }\Delta _{\varepsilon } &=&-\varepsilon k\left[ 1-\left(
ik_{0}-\varepsilon k\right) \Delta _{\varepsilon }\right] \,;  \nonumber \\
a_{\varepsilon }^{2}\Delta _{\varepsilon } &=&-\varepsilon k\left[
a_{\varepsilon }+\varepsilon k\left( ik_{0}-\varepsilon k\right) \left[
1-\left( ik_{0}-\varepsilon k\right) \Delta _{\varepsilon }\right] \right]
\,.  \label{linearize-delta-delta-etc}
\end{eqnarray}
After rearrangements, many terms happen to be real. Dropping these will
yield: 
\begin{eqnarray}
\mathop{\rm Im}%
\,^{\ast }\Sigma _{\varepsilon l}(P) &=&\frac{2}{\pi ^{2}T^{2}}%
\mathop{\rm Im}%
\,T\widetilde{\sum_{k_{0}}}\int_{\eta }^{+\infty }dk\,k^{2}\left[ \gamma
^{0}\left( \left( 1+\frac{ik_{0}-\varepsilon k}{ip_{0}}\right) ^{2}\Delta
_{\varepsilon }+\frac{\varepsilon }{p_{0}^{2}k}a_{\varepsilon }^{\prime
}\right) \right.  \nonumber \\
&&\hspace{-0.85in}-i\gamma ^{3}\frac{p}{3}\left[ 2\frac{ik_{0}-\varepsilon k%
}{p_{0}^{2}}\left( 1+\frac{ik_{0}-\varepsilon k}{ip_{0}}\right) \Delta
_{\varepsilon }-\varepsilon \left( 1+\frac{ik_{0}-\varepsilon k}{ip_{0}}%
\right) ^{2}\Delta _{\varepsilon }\partial _{k}-\frac{1}{p_{0}^{2}k}%
a_{\varepsilon }^{\prime }\left( \frac{2\varepsilon }{ip_{0}}+\partial
_{k}\right) \right]  \nonumber \\
&&\hspace{-0.85in}+\gamma ^{0}\frac{p^{2}}{3}\left[ \left( \frac{3}{%
2p_{0}^{4}k^{2}}-2\frac{ik_{0}-\varepsilon k}{ip_{0}^{3}}+\frac{\left(
ik_{0}-\varepsilon k\right) ^{2}}{2p_{0}^{4}}\left( 5+2\varepsilon \frac{%
ik_{0}}{k}-3\frac{k_{0}^{2}}{k^{2}}\right) \right. \right.  \nonumber \\
&&\hspace{-0.85in}-\left. \varepsilon \frac{ik_{0}-\varepsilon k}{p_{0}^{4}k}%
\left( 1+3\varepsilon \frac{ik_{0}}{k}\right) \right) \Delta _{\varepsilon
}+\left( \frac{1}{k}-\frac{2}{p_{0}^{2}k}+2\frac{ik_{0}\left(
ik_{0}-\varepsilon k\right) }{p_{0}^{2}k}+2\frac{ik_{0}-\varepsilon k}{%
ip_{0}k}\right.  \nonumber \\
&&\hspace{-0.85in}+\left. 2\frac{ik_{0}\left( ik_{0}-\varepsilon k\right)
^{2}}{ip_{0}^{3}k}-2\frac{ik_{0}-\varepsilon k}{ip_{0}^{3}k}-\frac{\left(
ik_{0}-\varepsilon k\right) ^{2}}{p_{0}^{2}k}\right) \Delta _{\varepsilon
}\partial _{k}+\frac{1}{2}\left( 1+\frac{ik_{0}-\varepsilon k}{ip_{0}}%
\right) ^{2}\Delta _{\varepsilon }\partial _{k}^{2}  \nonumber \\
&&\hspace{-0.85in}\left. \left. \hspace{-3pt}-\frac{\varepsilon }{2p_{0}^{4}k%
}\left( \hspace{-3pt}5+2\varepsilon \frac{ik_{0}}{k}-3\frac{k_{0}^{2}}{k^{2}}%
\right) a_{\varepsilon }^{\prime }-\hspace{-3pt}\varepsilon \left( 2\frac{%
ik_{0}}{ip_{0}^{3}k^{2}}-\frac{1}{p_{0}^{2}k^{2}}\right) a_{\varepsilon
}^{\prime }\partial _{k}+\hspace{-3pt}\frac{\varepsilon }{2p_{0}^{2}k}%
a_{\varepsilon }^{\prime }\partial _{k}^{2}\right] \hspace{-3pt}+\dots 
\hspace{-3pt}\right] \,^{\ast }\Delta _{l}\left( q_{0},k\right) .
\label{Im-sigma-pm-l--1}
\end{eqnarray}
Here $a_{\varepsilon }^{\prime }=a_{\varepsilon }-1=\varepsilon \left(
1-\varepsilon \frac{ik_{0}}{k}\right) Q_{0k}$. Since $ik_{0}$ appears in (%
\ref{Im-sigma-pm-l--1}) only in the numerator of fractions, we can sum over
it using the spectral decompositions (\ref{spectral-relations}). At each
time, we are left with two frequency integrals together with the one over $k$%
. Now we are allowed to take the real-energy analytic continuation $%
ip_{0}\rightarrow \omega _{\pm }(p)+i0^{+}$. But just before, every $e^{%
\frac{ip_{0}}{T}}$ has to be replaced with $-1$ except in the energy
denominators which occur only once in each term, thanks to the
rearrangements we made using (\ref{linearize-delta-delta-etc}). The
extraction of the imaginary part becomes straightforward if we use the
relation $1/\left( x+i0^{+}\right) =\Pr \left( 1/x\right) -i\pi \delta (x)$.
We obtain the following expression: 
\begin{eqnarray}
\mathop{\rm Im}%
\,^{\ast }\Sigma _{\varepsilon l}(P) &=&\frac{2}{\pi T}\int_{\eta }^{+\infty
}dk\,\int_{-\infty }^{+\infty }d\omega \int_{-\infty }^{+\infty }\frac{%
d\omega ^{\prime }}{\omega ^{\prime }}\delta \left( \omega _{\pm }-\omega
-\omega ^{\prime }\right) \left[ \gamma ^{0}\left( k^{2}\left( 1+\omega
-\varepsilon k\right) ^{2}\rho _{\varepsilon }\right. \right.  \nonumber \\
&&\hspace{-0.92in}-\left. \left( k-\varepsilon \omega \right) \rho
_{0}\right) +\frac{p}{3}\left[ \mp 2\gamma ^{0}\left( \omega -\varepsilon
k\right) \left( k^{2}\left( 1+\omega -\varepsilon k\right) \rho
_{\varepsilon }+\varepsilon \rho _{0}\right) \right.  \nonumber \\
&&\hspace{-0.92in}+\left. i\gamma ^{3}\left( 2k^{2}\left( \omega
-\varepsilon k\right) \left( 1+\omega -\varepsilon k\right) \rho
_{\varepsilon }+\varepsilon k^{2}\left( 1+\omega -\varepsilon k\right)
^{2}\rho _{\varepsilon }\partial _{k}+\left( \omega -\varepsilon k\right)
\rho _{0}\left( 2\varepsilon +\partial _{k}\right) \right) \right]  \nonumber
\\
&&\hspace{-0.92in}+\frac{p^{2}}{3}\left[ \gamma ^{0}\left[ \left( \frac{3}{2}%
-\varepsilon \left( \omega -\varepsilon k\right) \left( k+3\varepsilon
\omega \right) +\frac{2}{3}k^{2}\left( \omega -\varepsilon k\right) +\hspace{%
-0.03in}\frac{1}{2}\left( \omega -\varepsilon k\right) ^{2}\left( \hspace{%
-0.03in}3k^{2}+2\varepsilon \omega k+3\omega ^{2}\right) \right) \rho
_{\varepsilon }\right. \right.  \nonumber \\
&&\hspace{-0.92in}+k\left( 3+4\left( \omega -\varepsilon k\right)
+k^{2}-\omega ^{2}-2\omega \left( \omega -\varepsilon k\right) ^{2}\right)
\rho _{\varepsilon }\partial _{k}+\frac{k^{2}}{2}\left( 1+\omega
-\varepsilon k\right) ^{2}\rho _{\varepsilon }\partial _{k}^{2}  \nonumber \\
&&\hspace{-0.92in}+\left. \frac{\varepsilon }{2k^{2}}\left( \omega
-\varepsilon k\right) \left( 3k^{2}+2\varepsilon \omega k+3\omega
^{2}\right) \rho _{0}-\frac{\varepsilon }{k}\left( \omega -\varepsilon
k\right) \left( 2\omega -1\right) \rho _{0}\partial _{k}+\frac{\varepsilon }{%
2}\left( \omega -\varepsilon k\right) \rho _{0}\partial _{k}^{2}\right] 
\nonumber \\
&&\hspace{-0.92in}\mp \frac{1}{3}i\gamma ^{3}\left[ 2k^{2}\left( \omega
-\varepsilon k\right) \left( 2+3\omega -3\varepsilon k\right) \rho
_{\varepsilon }+2\varepsilon k^{2}\left( 1+\omega -\varepsilon k\right)
\left( \omega -\varepsilon k\right) \rho _{\varepsilon }\partial _{k}\right.
\nonumber \\
&&\hspace{-0.92in}\left. +\left. \left. 2\left( \omega -\varepsilon k\right)
\rho _{0}\left( 3\varepsilon +\partial _{k}\right) \right] \right] +\dots \,%
\right] \rho _{l}^{\prime }\,.  \label{Im-sigma-pm-l--2}
\end{eqnarray}
Recall that we have set $m_{f}=1$. The notation is as follows: $\rho
_{\varepsilon ,0}=\rho _{\varepsilon ,0}(\omega ,k)$; $\rho _{l}^{\prime
}=\rho _{l}(\omega ^{\prime },k)$. In the above expression, we have used $%
\tilde{n}(\omega )\simeq \frac{1}{2}$ and $n(\omega )\simeq T/\omega $. This
is because only soft values of $\omega $ and $\omega ^{\prime }$ are to
contribute. The resulting integrals are to be performed numerically, but
after the extraction of potential infrared divergences.

It remains to calculate the two other contributions to $%
\mathop{\rm Im}%
^{\ast }\Sigma _{1}$, those coming from transverse gluons in (\ref
{sigma-star-1}). The final result is the following: 
\begin{eqnarray}
\mathop{\rm Im}%
\,^{\ast }\Sigma _{\varepsilon t}(P) &=&\frac{2}{\pi T}\int_{\eta }^{+\infty
}dk\,\int_{-\infty }^{+\infty }d\omega \int_{-\infty }^{+\infty }\frac{%
d\omega ^{\prime }}{\omega ^{\prime }}\delta \left( \omega _{\pm }-\omega
-\omega ^{\prime }\right) \left[ \gamma ^{0}\left[ \left( -\frac{1}{2}%
(2k+\varepsilon )^{2}\right. \right. \right.  \nonumber \\
&&\hspace{-0.92in}-\left. \left. \frac{1}{2}\left( k^{2}-\omega ^{2}\right)
^{2}-\varepsilon \left( 2k+\varepsilon \right) \left( k^{2}-\omega
^{2}\right) \right) \rho _{\varepsilon }+\frac{1}{2k^{2}}\left( \varepsilon
\omega +k\right) \left( k^{2}-\omega ^{2}\right) \rho _{0}\right]  \nonumber
\\
&&\hspace{-0.92in}+\frac{p}{3}\left[ \mp \gamma ^{0}\left[ -\left( \left(
k^{2}-\omega ^{2}\right) ^{2}+2\left( 1+\varepsilon k\right) \left(
k^{2}-\omega ^{2}\right) +2\varepsilon k+1\right) \rho _{\varepsilon
}\right. \right.  \nonumber \\
&&\hspace{-0.92in}+\left. \frac{1}{k^{2}}\left( k+\varepsilon \omega \right)
\left( k^{2}-\omega ^{2}\right) \rho _{0}\right] +i\gamma ^{3}\left[ \left(
-\left( k^{2}-\omega ^{2}\right) ^{2}-\frac{2\varepsilon }{k}\left(
k^{2}-\omega ^{2}\right) \left( k^{2}+\varepsilon \omega k-\omega
^{2}\right) \right. \right.  \nonumber \\
&&\hspace{-0.92in}+\left. 4\omega (\omega -\varepsilon k)+4\varepsilon \frac{%
\omega ^{2}}{k}-2\omega -3-\frac{2\varepsilon }{k}\right) \rho _{\varepsilon
}+\varepsilon \left( \frac{1}{2}\left( k^{2}-\omega ^{2}\right)
^{2}+2\varepsilon k\left( k^{2}-\omega ^{2}\right) +3k^{2}-\omega ^{2}\right.
\nonumber \\
&&\hspace{-0.92in}\left. +\left. \left. 2\varepsilon k+\frac{1}{2}\right)
\rho _{\varepsilon }\partial _{k}+\left( 2\frac{\omega }{k^{3}}+\frac{1}{%
k^{2}}\left( k+\varepsilon \omega \right) \right) \left( k^{2}-\omega
^{2}\right) \rho _{0}-\frac{\varepsilon }{2k^{2}}\left( k+\varepsilon \omega
\right) ^{2}\left( k-\varepsilon \omega \right) \rho _{0}\partial _{k}\right]
\right]  \nonumber \\
&&\hspace{-0.92in}+\frac{p^{2}}{3}\hspace{-3pt}\left[ \gamma ^{0}\hspace{-1pt%
}\left[ \hspace{-0.03in}\hspace{-1pt}\left( \hspace{-0.03in}-k^{2}\left(
\omega -\varepsilon k\right) ^{2}\left( 1+\varepsilon \frac{\omega }{k}%
\right) \hspace{-0.03in}\hspace{-3pt}\left( \hspace{-3pt}1+\varepsilon \frac{%
\omega ^{3}}{k^{3}}\right) +\hspace{-0.03in}k\left( \omega -\varepsilon
k\right) \hspace{-0.03in}\left( \hspace{-0.03in}\frac{2}{3}k+\frac{%
11\varepsilon }{3}\omega -\frac{\omega ^{2}}{k}-\hspace{-0.03in}3\varepsilon 
\frac{\omega ^{3}}{k^{2}}+\hspace{-0.03in}\frac{\omega ^{4}}{k^{3}}\right)
\right. \right. \right.  \nonumber \\
&&\hspace{-0.92in}-\frac{k}{2}\left( \omega -\varepsilon k\right) \left( 
\hspace{-0.03in}\frac{7\varepsilon }{3}-\hspace{-0.03in}\frac{13\omega }{3k}-%
\frac{47\varepsilon }{3k^{2}}\omega ^{2}-\hspace{-0.03in}\frac{\omega ^{3}}{%
k^{3}}\right) -k\left( \hspace{-0.03in}\frac{4\varepsilon }{3}-\hspace{%
-0.03in}\frac{10\omega }{3k}+\hspace{-0.03in}\frac{2\varepsilon }{3k^{2}}%
\omega ^{2}+\hspace{-0.03in}\frac{2\omega ^{3}}{k^{3}}\right) -\hspace{%
-0.03in}\frac{4}{9}-\hspace{-0.03in}\frac{25\varepsilon }{3k}\omega +\hspace{%
-0.03in}2\frac{\omega ^{2}}{k^{2}}  \nonumber \\
&&\hspace{-0.92in}+\left. \frac{14\varepsilon }{3k}+\frac{\omega }{k^{2}}-%
\frac{3}{2k^{2}}\right) \rho _{\varepsilon }+(\frac{\omega }{k}\left(
k^{2}-\omega ^{2}\right) ^{2}-\frac{1}{2k}\left( k^{2}-\omega ^{2}\right)
\left( k^{2}-\omega ^{2}-4\varepsilon k\omega \right)  \nonumber \\
&&\hspace{-0.92in}+\left. \frac{1}{k}\left( \omega -\varepsilon k\right)
\left( \frac{4}{3}k^{2}-\frac{2\varepsilon }{3}k\omega -2\omega ^{2}\right) -%
\frac{5}{3}k+2\varepsilon \omega +\frac{\omega ^{2}}{k}-\frac{4\varepsilon }{%
3}+\frac{\omega }{k}-\frac{1}{2k}\right) \rho _{\varepsilon }\partial _{k} 
\nonumber \\
&&\hspace{-0.92in}+\frac{1}{4}\left( \hspace{-0.03in}-\left( \omega
^{2}-k^{2}\right) ^{2}+\hspace{-0.03in}4\varepsilon k\left( \omega
^{2}-k^{2}\right) -6k^{2}-4\varepsilon k+2\omega ^{2}-1\right) \rho
_{\varepsilon }\partial _{k}^{2}-\varepsilon \left( \hspace{-0.03in}\left(
1+\varepsilon \frac{\omega }{k}\right) \hspace{-0.03in}\left( 1+\varepsilon 
\frac{\omega ^{3}}{k^{3}}\right) \right.  \nonumber \\
&&\hspace{-0.92in}+\left. \frac{\omega }{k^{2}}\left( 3+2\varepsilon \frac{%
\omega }{k}-\frac{\omega ^{2}}{k^{2}}\right) -\frac{1}{2k^{2}}\left(
1+2\varepsilon \frac{\omega }{k}-3\frac{\omega ^{2}}{k^{2}}\right) \right)
\left( \omega -\varepsilon k\right) \rho _{0}-\varepsilon \left( \frac{1}{%
2k^{3}}-\frac{\omega }{k^{3}}\right)  \nonumber \\
&&\hspace{-0.92in}\left. \times \left( \omega ^{2}-k^{2}\right) \left(
\omega +\varepsilon k\right) \rho _{0}\partial _{k}+\frac{1}{4k^{2}}\left(
k+\varepsilon \omega \right) ^{2}\left( k-\varepsilon \omega \right) \rho
_{0}\partial _{k}^{2}\right]  \nonumber \\
&&\hspace{-0.92in}\mp \frac{1}{3}i\gamma ^{3}[(-3\left( k^{2}-\omega
^{2}\right) ^{2}-\hspace{-0.03in}\frac{4\varepsilon }{k}\left( k^{2}-\omega
^{2}\right) \left( k^{2}+\varepsilon k\omega -\omega ^{2}\right) +\hspace{%
-0.03in}4\left( \omega -\varepsilon k\right) \left( 2\omega +\varepsilon
k\right)  \nonumber \\
&&\hspace{-0.92in}-\left. 4\left( 2\varepsilon k+\omega -\hspace{-0.03in}%
2\varepsilon \frac{\omega ^{2}}{k}\right) -5-\frac{4\varepsilon }{k}\right)
\rho _{\varepsilon }+\left( \varepsilon \left( k^{2}-\omega ^{2}\right)
^{2}+2\left( k+\varepsilon \right) \left( k^{2}-\omega ^{2}\right)
+2k+\varepsilon \right) \rho _{\varepsilon }\partial _{k}  \nonumber \\
&&\hspace{-0.92in}\left. +\left. \left. \left( \frac{3}{k^{2}}\left(
k+\varepsilon \omega \right) +4\frac{\omega }{k^{3}}\right) \left(
k^{2}-\omega ^{2}\right) \rho _{0}+\frac{1}{k^{2}}\left( \omega -\varepsilon
k\right) \left( k+\varepsilon \omega \right) ^{2}\rho _{0}\partial _{k}%
\right] \right] +\dots \right] \rho _{t}^{\prime }\,.  \label{Im-sigma-pm-t}
\end{eqnarray}
This expression is quite long because these two terms are more involved.
However, there are not new steps worth discussing in detail.

\subsection{Calculation of $%
\mathop{\rm Im}%
\,^{\ast }\Sigma _{2}\left( P\right) $}

Now we turn to calculating the imaginary part of $^{\ast }\Sigma _{2}\left(
P\right) $ which can be written from (\ref{eff-self-energy2}) and the
structure of the gluon propagator in the strict Coulomb gauge as: 
\begin{equation}
^{\ast }\Sigma _{2}\left( P\right) =-\frac{8}{T^{2}}\,T\sum_{k_{0}}\int 
\frac{d^{3}k}{(2\pi )^{3}}\int \frac{d\Omega _{s}}{4\pi }\frac{iS\hspace{%
-0.6206pc}/}{KS\,PS\,QS}\left[ \,^{\ast }\Delta _{l}(K)-\left( 1-\widehat{%
{\bf k}}.\widehat{{\bf s}}^{2}\right) \,^{\ast }\Delta _{t}(K)\right] .
\label{sigma2-1}
\end{equation}
The steps to carry are similar to the ones we used for the previous
contribution. But here we need the additional expansion: 
\begin{equation}
\frac{1}{QS}=\frac{1}{iq_{0}-\widehat{{\bf k}}.\widehat{{\bf s}}}\left[ 1-%
\frac{\widehat{{\bf p}}.\widehat{{\bf s}}}{iq_{0}-\widehat{{\bf k}}.\widehat{%
{\bf s}}}-\frac{\widehat{{\bf p}}.\widehat{{\bf s}}^{2}}{\left( iq_{0}-%
\widehat{{\bf k}}.\widehat{{\bf s}}\right) ^{2}}+\dots \right] \,.
\label{expansion-1/QS}
\end{equation}
The angular integrals over the solid angles $\Omega _{s}$ and then $\Omega
_{k}$ are done as usual. We obtain for the longitudinal gluon: 
\begin{eqnarray}
^{\ast }\Sigma _{2l}\left( P\right) &=&-\frac{4}{\pi ^{2}T^{2}}%
\,T\sum_{k_{0}}\int_{\eta }^{+\infty }k\,dk\left[ \gamma ^{0}\frac{1}{%
p_{0}^{2}}Q_{0k}^{\prime }+i\gamma ^{3}\frac{p}{3}\left[ \frac{2}{ip_{0}^{3}}%
Q_{0k}^{\prime }-\frac{k}{p_{0}^{2}\left( q_{0}^{2}+k^{2}\right) }\right]
\right.  \nonumber \\
&&+\left. \gamma ^{0}\frac{p^{2}}{p_{0}^{2}}\left[ -\frac{1}{p_{0}^{2}}%
Q_{0k}^{\prime }-\frac{2k}{3ip_{0}\left( q_{0}^{2}+k^{2}\right) }+\frac{%
iq_{0}k}{3\left( q_{0}^{2}+k^{2}\right) ^{2}}\right] +\dots \right] \,^{\ast
}\Delta _{l}(k_{0},k)\,,  \label{sigma2l-intermediary1}
\end{eqnarray}
and for the transverse one: 
\begin{eqnarray}
^{\ast }\Sigma _{2t}\left( P\right) &=&-\frac{4}{\pi ^{2}T^{2}}%
\,T\sum_{k_{0}}\int_{\eta }^{+\infty }\frac{dk}{k}\left[ -\gamma ^{0}\frac{1%
}{p_{0}^{2}}\left( q_{0}^{2}+k^{2}\right) Q_{0k}^{\prime }-i\gamma ^{3}\frac{%
2p}{3p_{0}^{2}}\left( \frac{1}{ip_{0}}\left( q_{0}^{2}+k^{2}\right)
+iq_{0}\right) Q_{0k}^{\prime }\right.  \nonumber \\
&&+\left. \gamma ^{0}\frac{p^{2}}{p_{0}^{2}}\left[ \left( \frac{1}{p_{0}^{2}}%
\left( q_{0}^{2}+k^{2}\right) -\frac{4iq_{0}}{3ip_{0}}+\frac{1}{3}\right)
Q_{0k}^{\prime }+\frac{iq_{0}k}{3\left( q_{0}^{2}+k^{2}\right) }\right]
+\dots \right] \,^{\ast }\Delta _{t}(k_{0},k)\,,
\label{sigma2t-intermediary1}
\end{eqnarray}
where $Q_{0k}^{\prime }$stands for $Q_{0}\left( \frac{iq_{0}}{k}\right) $.
The sum over bosonic $k_{0}$ is now readily done if we add to the spectral
representations (\ref{spectral-relations}) those of $1/\left(
q_{0}^{2}+k^{2}\right) $ and $1/\left( q_{0}^{2}+k^{2}\right) ^{2}$. We
have: 
\begin{eqnarray}
\frac{1}{\left( q_{0}^{2}+k^{2}\right) } &=&\int_{0}^{1/T}d\tau
\,e^{iq_{0}\tau }\int_{-\infty }^{+\infty }d\omega \,\epsilon (\omega
)\delta (\omega ^{2}-k^{2})\left( 1-\tilde{n}(\omega )\right) e^{-\omega
\tau }\,;  \nonumber \\
\frac{1}{\left( q_{0}^{2}+k^{2}\right) ^{2}} &=&\int_{0}^{1/T}d\tau
\,e^{iq_{0}\tau }\int_{-\infty }^{+\infty }d\omega \,\epsilon (\omega
)\,\delta ^{(1)}(\omega ^{2}-k^{2})\left( 1-\tilde{n}(\omega )\right)
e^{-\omega \tau }\,,  \label{spectral-relations-1/k**2}
\end{eqnarray}
with $q_{0}$ fermionic. $\epsilon (\omega )$ is the sign function and $%
\delta ^{(1)}(\omega ^{2}-k^{2})$ stands for $\partial _{\omega ^{2}}\delta
\left( \omega ^{2}-k^{2}\right) $. The extraction of the imaginary part is
straightforward. We obtain for the longitudinal contribution: 
\begin{eqnarray}
\mathop{\rm Im}%
\,^{\ast }\Sigma _{2l}(P) &=&-\frac{4}{\pi T}\int_{\eta }^{+\infty
}dk\int_{-\infty }^{+\infty }d\omega \int_{-\infty }^{+\infty }\frac{d\omega
^{\prime }}{\omega ^{\prime }}\delta (\omega _{\pm }-\omega -\omega ^{\prime
})\left[ -\gamma ^{0}k\rho _{0}+\frac{p}{3}\left[ \pm 2\gamma ^{0}k\rho
_{0}\right. \right.  \nonumber \\
&&\hspace{-0.4in}+\left. i\gamma ^{3}\left( -2k\rho _{0}+k^{2}\epsilon
(\omega )\delta \left( \omega ^{2}-k^{2}\right) \right) \right] +p^{2}\left[
\gamma ^{0}\left( -\frac{2k}{3}\rho _{0}+\frac{2k^{2}}{3}\epsilon (\omega
)\delta (\omega ^{2}-k^{2})\right. \right.  \nonumber \\
&&\hspace{-0.4in}\left. -\left. \left. \frac{k^{2}}{3}\omega \epsilon
(\omega )\,\delta ^{(1)}(\omega ^{2}-k^{2})\right) \mp \frac{1}{3}i\gamma
^{3}\left( -2k\rho _{0}+\frac{2k^{2}}{3}\epsilon (\omega )\delta \left(
\omega ^{2}-k^{2}\right) \right) \right] +\dots \right] \,\rho _{l}^{\prime
}\,,  \label{Im-sigma-2l-final}
\end{eqnarray}
and for the transverse one: 
\begin{eqnarray}
\mathop{\rm Im}%
\,^{\ast }\Sigma _{2t}(P) &=&-\frac{4}{\pi T}\int_{\eta }^{+\infty
}dk\int_{-\infty }^{+\infty }d\omega \int_{-\infty }^{+\infty }\frac{d\omega
^{\prime }}{\omega ^{\prime }}\delta (\omega _{\pm }-\omega -\omega ^{\prime
})\left[ \gamma ^{0}\frac{1}{k}\left( k^{2}-\omega ^{2}\right) \rho
_{0}\right.  \nonumber \\
&&\hspace{-0.9in}+\frac{p}{3}\left[ \mp \gamma ^{0}\frac{2}{k}\left(
k^{2}-\omega ^{2}\right) \rho _{0}+i\gamma ^{3}\frac{2}{k}\left(
k^{2}+\omega -\omega ^{2}\right) \rho _{0}\right] +p^{2}\left[ \gamma
^{0}\left( \frac{2}{3k}\left( k^{2}-\frac{1}{2}+2\omega -\omega ^{2}\right)
\rho _{0}\right. \right.  \nonumber \\
&&\hspace{-0.9in}\left. \left. \left. -\frac{\omega }{3}\epsilon (\omega
)\delta \left( \omega ^{2}-k^{2}\right) \right) \mp i\gamma ^{3}\frac{2}{3k}%
\left( k^{2}+\frac{2}{3}\omega -\omega ^{2}\right) \rho _{0}\right] +\dots %
\right] \,\rho _{t}^{\prime }\,.  \label{Im-sigma2t-final}
\end{eqnarray}

The final result we aim at is the sum of the six terms: 
\begin{equation}
\mathop{\rm Im}%
\,^{\ast }\Sigma (P)=\sum_{\varepsilon =\pm ,\,i=l,t}%
\mathop{\rm Im}%
\,^{\ast }\Sigma _{\varepsilon i}(P)+\sum_{i=l,t}%
\mathop{\rm Im}%
\,^{\ast }\Sigma _{2i}(P)\,,  \label{Im-sigma-final}
\end{equation}
where the different contributions are given in (\ref{Im-sigma-pm-l--2}), (%
\ref{Im-sigma-pm-t}), (\ref{Im-sigma-2l-final}) and (\ref{Im-sigma2t-final}%
). To get the damping rates $\gamma _{\pm }\left( p\right) $, we use eq (\ref
{gamma-pm-b}) where $^{\ast }f_{\pm }=\,^{\ast }D_{0}\mp \,^{\ast }D_{s}$
and $^{\ast }\Sigma =\gamma ^{0}\,^{\ast }D_{0}+i{\bf \gamma }.\widehat{{\bf %
p}}\,\,^{\ast }D_{s}$. There are few more steps though. Indeed, note that
the energy $\omega _{\pm }(p)$ appearing in $\delta (\omega _{\pm }-\omega
-\omega ^{\prime })$ is a function of $p$, given in (\ref{omega-pm}) for
small $p$. This means that for the terms in $p^{2}$, the energy $\omega
_{\pm }(p)$ can be replaced by one (in units of $m_{f})$ since we look for
the damping rates up to order $p^{2}$, but for the terms of order $p$, we
have to expand $\delta (\omega _{\pm }-\omega -\omega ^{\prime })$ to order $%
p$ and for the terms of order zero to order $p^{2}$. A subsequent
rearrangement is necessary.

\section{Results and conclusion \ \ \ }

The damping rates are given in (\ref{gamma-pm-b}). We find: 
\begin{equation}
\gamma _{\pm }\left( p\right) =-\frac{g^{2}C_{f}T}{8\pi }\left[ a_{0}\pm 
\frac{p}{3}a_{1}+\frac{p^{2}}{9}a_{2}+\dots\right] ,  \label{gam-pm-final}
\end{equation}
where the coefficients $a_{i}$ are given by the expressions: 
\begin{eqnarray}
a_{0} &=&\int_{\eta }^{\infty }dk\int_{-\infty }^{+\infty }d\omega
\int_{-\infty }^{+\infty }\frac{d\omega ^{\prime }}{\omega ^{\prime }}%
f_{0}\left( \omega ,\omega ^{\prime };k\right) \delta \,;  \nonumber \\
a_{1} &=&\int_{\eta }^{\infty }dk\int_{-\infty }^{+\infty }d\omega
\int_{-\infty }^{+\infty }\frac{d\omega ^{\prime }}{\omega ^{\prime }}\left[
\,f_{1}\left( \omega ,\omega ^{\prime };k\right) -\,f_{0}\left( \omega
,\omega ^{\prime };k\right) \,\partial _{\omega }\right] \delta ,  \nonumber
\\
a_{2} &=&\int_{\eta }^{\infty }\hspace{-3pt}dk\int_{-\infty }^{+\infty }%
\hspace{-3pt}d\omega \int_{-\infty }^{+\infty }\frac{d\omega ^{\prime }}{%
\omega ^{\prime }}\hspace{-2pt}\left[ \hspace{-3pt}f_{2}\left( \omega
,\omega ^{\prime };k\right) -\hspace{-2pt}f_{1}\left( \omega ,\omega
^{\prime };k\right) \partial _{\omega }+\hspace{-2pt}f_{0}\left( \omega
,\omega ^{\prime };k\right) \left[ -3\,\partial _{\omega }+\partial _{\omega
}^{2}\right] \right] \delta ,  \label{a0-a1-a2}
\end{eqnarray}
with $\delta =\delta \left( 1-\omega -\omega ^{\prime }\right) $. The three
functions $f_{i}\left( \omega ,\omega ^{\prime };k\right) $ are given by the
following expressions:

\begin{eqnarray}
f_{0}\left( \omega ,\omega ^{\prime };k\right) &=&\sum_{\varepsilon =\pm } 
\left[ -k^{2}\left( 1-\varepsilon k+\omega \right) ^{2}\rho _{\varepsilon
}\rho _{l}^{\prime }+\frac{1}{2}\left( 1+2\varepsilon k+k^{2}-\omega
^{2}\right) ^{2}\rho _{\varepsilon }\rho _{t}^{\prime }\right]  \nonumber \\
&&+\frac{1}{k}\left( k^{2}-\omega ^{2}\right) \rho _{0}\rho _{t}^{\prime }\,.
\label{f0}
\end{eqnarray}
This expression is the one obtained in \cite
{kobes-kunstatter-mak,braaten-pisarski (quarks)}. $f_{1}$ and $f_{2}$ are
new. They read: 
\begin{eqnarray}
f_{1}\left( \omega ,\omega ^{\prime };k\right) &=&\sum_{\varepsilon =\pm } 
\left[ 2k^{2}\left( -1+k^{2}-2\varepsilon k\omega +\omega ^{2}\right) \rho
_{\varepsilon }\rho _{l}^{\prime }+\left( -\frac{2\varepsilon }{k}%
-3+2\varepsilon k+4k^{2}-k^{4}\right. \right.  \nonumber \\
&&\ \hspace{-1.1in}\left. -\left( 2+4\varepsilon k+2k^{2}\right) \omega
+\left( \frac{4\varepsilon }{k}+4+2\varepsilon k+2k^{2}\right) \omega
^{2}+2\omega ^{3}-\left( \frac{2\varepsilon }{k}+1\right) \omega ^{4}\right]
\rho _{\varepsilon }\rho _{t}^{\prime }  \nonumber \\
&&\ \ \hspace{-1.1in}+\left. \hspace{-3pt}\varepsilon k^{2}\hspace{-2pt}%
\left( 1\hspace{-2pt}-\varepsilon k+\hspace{-2pt}\omega \right) ^{2}\hspace{%
-3pt}\rho _{\varepsilon }\partial _{k}\rho _{l}^{\prime }+\hspace{-1pt}%
\left( \frac{\varepsilon }{2}+2k+3\varepsilon k^{2}+2k^{3}+\frac{\varepsilon 
}{2}k^{4}-\left( \varepsilon +2k+\varepsilon k^{2}\right) \allowbreak \omega
^{2}+\frac{\varepsilon }{2}\omega ^{4}\allowbreak \right) \hspace{-3pt}\rho
_{\varepsilon }\partial _{k}\rho _{t}^{\prime }\right]  \nonumber \\
&&\ \ \hspace{-1.1in}-\frac{2}{k}\left( k^{2}-\omega ^{2}+2\frac{\omega ^{3}%
}{k^{2}}\right) \hspace{-3pt}\rho _{0}\rho _{t}^{\prime }-\hspace{-3pt}%
2k^{2}\epsilon \left( \omega \right) \delta \left( \omega ^{2}-k^{2}\right) 
\hspace{-3pt}\rho _{l}^{\prime }+\hspace{-3pt}%
{\displaystyle{\omega  \over k^{2}}}%
\hspace{-3pt}\left( \omega ^{2}-k^{2}\right) \hspace{-3pt}\rho _{0}\partial
_{k}\rho _{t}^{\prime }+2\omega \rho _{0}\partial _{k}\rho _{l}^{\prime }\,;
\label{f1}
\end{eqnarray}
and: 
\begin{eqnarray}
f_{2}\left( \omega ,\omega ^{\prime };k\right) &=&\sum_{\varepsilon =\pm }%
\hspace{-3pt}\left[ \left( -\frac{9}{2}-k^{2}-6\varepsilon k^{3}-\frac{1}{2}%
k^{4}-\left( 6\varepsilon k-6k^{2}+2\varepsilon k^{3}\right) \allowbreak
\omega +\left( 9+k^{2}\right) \omega ^{2}+6\varepsilon k\omega ^{3}\right.
\allowbreak \right.  \nonumber \\
&&\hspace{-1.03in}-\left. \frac{9}{2}\omega ^{4}\hspace{-3pt}\right) \hspace{%
-3pt}\rho _{\varepsilon }\rho _{l}^{\prime }+\hspace{-3pt}\left( \frac{9}{%
2k^{2}}-\frac{14\varepsilon }{k}-\hspace{-2pt}\frac{8}{3}+\hspace{-3pt}%
4\varepsilon k-\hspace{-2pt}\frac{19}{2}k^{2}\hspace{-2pt}-\hspace{-3pt}%
6\varepsilon k^{3}\hspace{-2pt}+\hspace{-2pt}k^{4}\allowbreak \hspace{-2pt}+%
\hspace{-3pt}\left( \frac{-3}{k^{2}}+\frac{25\varepsilon }{k}\hspace{-2pt}-%
\hspace{-2pt}10\hspace{-2pt}+\hspace{-3pt}6\varepsilon k\hspace{-2pt}+%
\hspace{-2pt}9k^{2}\hspace{-2pt}-\hspace{-3pt}3\varepsilon k^{3}\hspace{-2pt}%
\right) \omega \right.  \nonumber \\
&&\hspace{-1.03in}+\left( -\frac{6}{k^{2}}+\frac{2\varepsilon }{k}%
+23-6\varepsilon k+k^{2}\right) \allowbreak \omega ^{2}+\left( \frac{6}{k^{2}%
}-\frac{22\varepsilon }{k}-6+6\varepsilon k\right) \allowbreak \omega
^{3}+\left( -\frac{3}{2k^{2}}+\frac{12\varepsilon }{k}-5\right) \allowbreak
\omega ^{4}  \nonumber \\
&&\hspace{-1.03in}-\left. \left( \frac{3}{k^{2}}+\frac{3\varepsilon }{k}%
\right) \allowbreak \omega ^{5}+\frac{3}{k^{2}}\allowbreak \omega
^{6}\right) \rho _{\varepsilon }\rho _{t}^{\prime }-k\left( 9-14\varepsilon
k+5k^{2}+\left( 12-2\varepsilon k-6k^{2}\right) \omega -\left(
3-12\varepsilon k\right) \omega ^{2}\right.  \nonumber \\
&&\hspace{-1.03in}-\left. 6\omega ^{3}\right) \rho _{\varepsilon }\partial
_{k}\rho _{l}^{\prime }+\left( \frac{3}{2k}+4\varepsilon +7k+8\varepsilon
k^{2}+\frac{7}{2}k^{3}-\left( \frac{3}{k}+6\varepsilon +6k+6\varepsilon
k^{2}+3k^{3}\right) \omega \right.  \nonumber \\
&&\hspace{-1.03in}-\left. \left( \frac{3}{k}+4\varepsilon +5k\right) \omega
^{2}+\left( \frac{6}{k}+6\varepsilon +6k\right) \omega ^{3}+\frac{3}{2k}%
\omega ^{4}-\frac{3}{k}\omega ^{5}\right) \rho _{\varepsilon }\partial
_{k}\rho _{t}^{\prime }-\frac{3}{2}k^{2}\left( 1-\varepsilon k+\omega
\right) ^{2}\rho _{\varepsilon }\partial _{k}^{2}\rho _{l}^{\prime } 
\nonumber \\
&&\hspace{-1.03in}+\left. \left( \frac{3}{4}+3\varepsilon k+\frac{9}{2}%
k^{2}+3\varepsilon k^{3}+\frac{3}{4}k^{4}-\frac{3}{2}\left( 1+2\varepsilon
k+k^{2}\right) \omega ^{2}+\frac{3}{4}\omega ^{4}\right) \rho _{\varepsilon
}\partial _{k}^{2}\rho _{t}^{\prime }\right] \hspace{-3pt}-\frac{3}{k}\left(
k^{2}-\omega ^{2}\right) \rho _{0}\rho _{l}^{\prime }  \nonumber \\
&&\hspace{-1.03in}+\left( \frac{3}{k}+2k+\frac{6}{k}\omega -\left( \frac{15}{%
k^{3}}+\frac{2}{k}\right) \omega ^{2}+\frac{18}{k^{3}}\omega ^{3}\right)
\rho _{0}\rho _{t}^{\prime }+\left( 6-12\omega \right) \rho _{0}\partial
_{k}\rho _{l}^{\prime }  \nonumber \\
&&\hspace{-1.03in}+\left( -3+6k\omega +\frac{3}{k^{2}}\omega ^{2}-\frac{6}{k}%
\omega ^{3}\right) \rho _{0}\partial _{k}\rho _{t}^{\prime }+3k\rho
_{0}\partial _{k}^{2}\rho _{l}^{\prime }-\frac{3}{2k}\left( k^{2}-\omega
^{2}\right) \rho _{0}\partial _{k}^{2}\rho _{t}^{\prime }  \nonumber \\
&&\hspace{-1.03in}+12k^{2}\epsilon \left( \omega \right) \delta \left(
\omega ^{2}-k^{2}\right) \rho _{l}^{\prime }-6\left| \omega \right| \delta
\left( \omega ^{2}-k^{2}\right) \rho _{t}^{\prime }-6k^{2}\left| \omega
\right| \,\partial _{\omega ^{2}}\delta \left( \omega ^{2}-k^{2}\right) \rho
_{l}^{\prime }\,.
\end{eqnarray}

It remains to perform the integrals over the frequencies $\omega $ and $%
\omega ^{\prime }$ and then over the momentum $k.$ Of course, these
integrations are not straightforward and necessitate numerical work \cite
{ABD}. Also, the dimensionless parameter $m(N_{c},N_{f})=m_{g}/m_{f}=\frac{4%
}{3}\sqrt{\frac{N_{c}(N_{c}+N_{f}/2)}{N_{c}^{2}-1}}$ is implicitly present
in the spectral densities $\rho _{l,t}$ and so, each case has to be treated
separately.

Recall that this direct calculation is performed in the sole context of the
Braaten-Pisarski HTL-summed next-to-leading order perturbation. As we
emphasized in the introductory remarks, there is the problem of occurrence
of infrared divergences to be aware of. Hence, extra work is needed in order
to extract these from the finite contributions. One interesting point to
wander about is what sort of divergences we will obtain. Indeed, in the
direct calculation of $\gamma _{l}(0)$, the damping rate for longitudinal
gluons with zero momentum, the divergent term behaves like $1/\eta ^{2}$ 
\cite{AA}. In the second coefficient in $p^{2}$ of $\gamma _{t}(p)$, the
damping rate for transverse gluons, $1/\eta ^{2}$ does also appear together
with $\ln \eta $ \cite{AABo}. The question is then: what sort of divergences
will we get for $\gamma _{\pm }(p)$? If different from $1/\eta ^{2}$ and $%
\ln \eta $, are we able to understand why?

We stress once again that the occurrence of these divergences may simply be
due to the early expansion we make of the HTL-summed next-to-leading order
self-energies in powers of the external momentum. We have argued otherwise
in section three, but a really more convincing argument would be to carry
the very same calculation in a way that avoids such an early expansion. We
have indicated that, in the perturbative context, this could be technically
very difficult.

In any case, there is by now convincing evidence in the literature that
next-to-leading order quantities are magnetic-sensitive. We tend to be of
the viewpoint that the occurrence of infrared divergences in HTL-summed
next-to-leading order self-energies is probably a manifestation of this
magnetic sensitivity, and that a more complete next-to-leading order
calculation should remove them. We mean a calculation that takes into
account magnetic effects not present at the electric scale, and hence not
incorporated in the hard thermal loops. Of course, we may need to understand
first such effects. It is then interesting to ask whether the infrared
divergences one obtains, in the damping rates and in possibly similar
quantities, can be of any help. It may also be that these effects cannot
even be grasped perturbatively.

Also, we have argued in section three that regularizing the infrared region
with a simple shift in the static transverse gluonic propagator by a
momentum-independent magnetic mass may not be the best way to shield from
magnetic sensitivity. This is important in view of the fact that much of the
results of high temperature QCD rely on such a regularization. In our work,
we exclude from the outset the magnetic region by introducing an infrared
cutoff of the order of the magnetic scale. One drawback is that our
calculation is valid only for very soft external momenta and the results we
obtain cannot be carried to larger values. That our results differ
analytically from other estimations, particularly those relying on the
magnetic-mass shielding, comes mainly from the regularization procedure and
its implications. Also, the difference in the results accentuates the
sensitivity of the HTL-based perturbation to the magnetic scale and is added
evidence that possibly interesting physics is happening between the soft and
very soft regions.

\end{document}